\documentclass[runningheads]{svmult}

\usepackage{makeidx}   
\usepackage{graphicx}  
\usepackage{subeqnar}  
\usepackage{multicol}  
\usepackage{physprbb}  
\makeindex             

\begin{document}
\input{epsf}
\title*{ The Two-State Vector Formalism: \protect\newline an Updated Review}

\toctitle{ The Two-State Vector Formalism: an Updated Review
\protect\newline }
%
%
\titlerunning{The Two-State Vector Formalism}
%
\author{Yakir Aharonov\inst{1,2}
\and Lev Vaidman\inst{1}}

\authorrunning{Yakir Aharonov and Lev Vaidman}
%
%
\institute{ School of Physics and Astronomy \\
Raymond and Beverly Sackler Faculty of Exact Sciences \\
Tel Aviv University, Tel-Aviv 69978, Israel \and Department of
Physics and Department of Computational and Data Sciences \\
College of Science, George Mason University, Fairfax, VA 22030}

\maketitle              

\begin{abstract}

  In this paper we present the two-state vector formalism of quantum
  mechanics. It is a time-symmetrized approach to standard quantum
  theory particularly helpful for the analysis of experiments
  performed on pre- and post-selected ensembles. Several peculiar effects
  which naturally arise in this approach are considered. In
  particular, the concept of ``weak measurements'' (standard
  measurements with weakening of the interaction) is  discussed in depth
  revealing a very unusual but consistent picture. Also, a design of
  a gedanken experiment which implements a kind of
  quantum ``time machine'' is described. The issue of time-symmetry in
  the context of the two-state vector formalism is clarified.

\end{abstract}

\section{Descriptions of quantum systems}
\label{Descr}

\subsection{The quantum state}
\label{sec:Intr}

In the standard quantum mechanics, a system at a given time $t$ is
described completely by a quantum state
\begin{equation}
  \label{1qs}
  |\Psi\rangle ,
\end{equation}
defined by the results of measurements performed on
the system in the past relative to the time $t$. (It might be that
the system at time $t$ is not described by a pure quantum state, but
by a mixed state (density matrix). However,  we can
always assume that there is a composite system  including this system
which is in a pure state.) The status of a quantum state is
controversial: there are many papers on reality of a quantum state
and numerous interpretations of this ``reality''. However, it is
 non-controversial to say  that the quantum state yields maximal
 information about how this system can affect other systems (in
 particular, measuring devices) interacting with it at time $t$. Of
 course, the results of all measurements in the past, or just the
 results of the last complete measurement, also have this information,
 but these results include other facts too,  so the quantum state is the most
 concise information about how the quantum system can affect other
 systems at time $t$.

 The concept of a quantum state is time-asymmetric: it is defined by
 the results of measurements in the {\it past}.  This fact by itself
 is not enough for the asymmetry: in classical physics, the state of a
 system at time $t$ defined by the results of the complete set of
 measurements in the past is not different from the state defined by
 the complete measurements in the future. This is because for
 a classical system the results of measurements in the future are
 defined by the results of measurements in the past (and vice versa).
 In quantum mechanics this is not so: the results of measurements in
 the future are only partially constrained by the results of measurements
 in the past. Thus, the concept of a quantum state is genuinely
 time-asymmetric. The question arises: does the asymmetry of a quantum
 state reflects the time asymmetry of quantum mechanics, or it can be
 removed by reformulation of quantum mechanics in a time-symmetric
 manner?

\subsection{The  two-state vector}
\label{sec:2sv}

The  two-state vector formalism of quantum mechanics (TSVF) originated
in a seminal work of Aharonov, Bergmann, and Lebowitz (ABL) \cite{ABL}
removes this asymmetry. It provides a time-symmetric formulation of
quantum mechanics.  A  system at a given time $t$ is described completely
by a {\it two-state vector}
\begin{equation}
  \label{2sv}
  \langle \Phi|~|\Psi\rangle ,
\end{equation}
which consists of a quantum state $|\Psi\rangle$ defined by the results
of measurements performed on the system in the past relative to the
time $t$ and of a backward evolving quantum state $ \langle \Phi|$ defined by the
results of measurements performed on this system after the  time $t$. Again, the status of the two-state vector might be
interpreted in different ways, but a non-controversial fact is
that it   yields maximal
 information about how this system can affect other systems (in
 particular, measuring devices) interacting with it at time $t$.

 The description of the system with the two-state vector
 (\ref{2sv}) is clearly different from the description with a
 single quantum state (\ref{1qs}), but in both cases we claim that
 ``it yields maximal information about how this system can affect
 other systems (in particular, measuring devices) interacting with it
 at time $t$.'' Does it mean that the TSVF has different predictions
 than the standard quantum approach? No, the two formalisms describe the
 same theory with the same predictions. The difference is that the
 standard approach is time asymmetric and it is assumed that only the
 results of the measurements in the past exist. With this constrain,
 $|\Psi\rangle$ indeed contains maximal information about the
 system at time $t$. The rational for this approach is that if the
 results of the future measurements relative to the time $t$ exist too,
 then ``now'' is after time $t$ and we cannot return back in time to
 perform measurements at $t$. Therefore, taking into account results of
 future measurements is not useful. In contrast, the TSVF approach is time
 symmetric. There is no preference to the results of measurements in
 the past relative to the results of measurements in the future: both
 are taken into account. Then, there is more information about the
 system at time $t$. The maximal information (without constrains) is
 contained in the two-state vector $ \langle \Phi|~|\Psi\rangle$.


  If the TSVF has the same predictions as standard quantum mechanics,
  what is the reason to consider it? And what about the argument that
 when the results of future measurements are known it is already too
  late to make measurements at time $t$? How the two-state vector might
  be useful? The answer to the first question is that it is important
  to understand the time-symmetry of nature (described by quantum
  mechanics).
  The time-asymmetry of the standard approach might be solely due to
 the usage of   time-asymmetric concepts. The answer to the second
  question is that there are many situations in which we want to know
  how a system affected other systems in the past. The TSVF proved to
  be particularly useful after introduction of {\it weak measurements}
  \cite{AACV,AV90,AV91} which allowed to see that systems described by
  some two-state vectors can affect other system at time $t$ in a very peculiar
  way.  This has led to the discovery of numerous bizarre effects
  \cite{s-xyz,s-100,t-m,K<0}. It is very difficult to understand
  these effects in  the framework of standard quantum mechanics;
  some of them can
  be explained via miraculous interference phenomenon known as  {\it
  super-oscillations} \cite{suos1,suos2}.

\subsection{How to create quantum systems corresponding to
various complete descriptions?}

The maximal complete description of a quantum system at time $t$ is a
two-state vector (\ref{2sv}). We will name the system which has such a
description as {\it pre- and post-selected}.  (Again, it might be that
at time $t$ the system is not described by a ``pure'' two-state
vector. However, we can assume that there is a composite system
including this system which is described by a two-state vector.)  In
some circumstances, the system might have only a partial description.
For example, if time $t$ is ``present'' and the results of the future
measurements do not exist yet, then at that time, the system is
described only by a usual forward evolving quantum state (\ref{1qs}):
the {\it pre-selected} system. Later, when the results of the future
measurements will be obtained, the description will be completed to
the form (\ref{2sv}). It is also possible to arrange a situation in
which, until some measurements in the future, the complete description
of the system at time $t$ is the backward evolving quantum state $
\langle \Phi|$: the {\it post-selected} system. We will now explain
how all these situations can be achieved.

\vskip .4cm
\noindent{\bf Single forward-evolving quantum state}

In order to have {\it now} a system the complete description of which
at time $t$ is a single quantum state (\ref{1qs}), there should be a
complete measurement in the past of time $t$ and no measurement on the
system after time $t$, see  Fig. 1{\it a}. The system in the state $|\Psi \rangle$ is obtained
when  a measurement of an
observable $A$ at time $t_1$ is performed, $t_1<t$, obtaining a specific outcome
$A=a$ such that the created state $|a\rangle$ performs unitary
evolution between $t_1$ and $t$ governed by the Hamiltonian $H$,
\begin{equation}
  \label{U}
 U(t_1, t) ~=~ e^{-i\int_{t_1}^t H dt}  ,
\end{equation}
to the desired
state:
\begin{equation}
  \label{ket=}
 |\Psi \rangle ~=~  U(t_1, t) ~|a\rangle .
\end{equation}
The time ``now'', $t_{now}$ should either be equal to the time $t$, or it
should be known that during the time period $[t, t_{now}]$ no
measurements have been performed on the system. The state
$|\Psi\rangle$  remains to be the complete description of the system at time $t$ until the future measurements on the system will be performed yielding additional information.

\vskip .5cm
\epsfxsize=12cm
 \centerline{\epsfbox{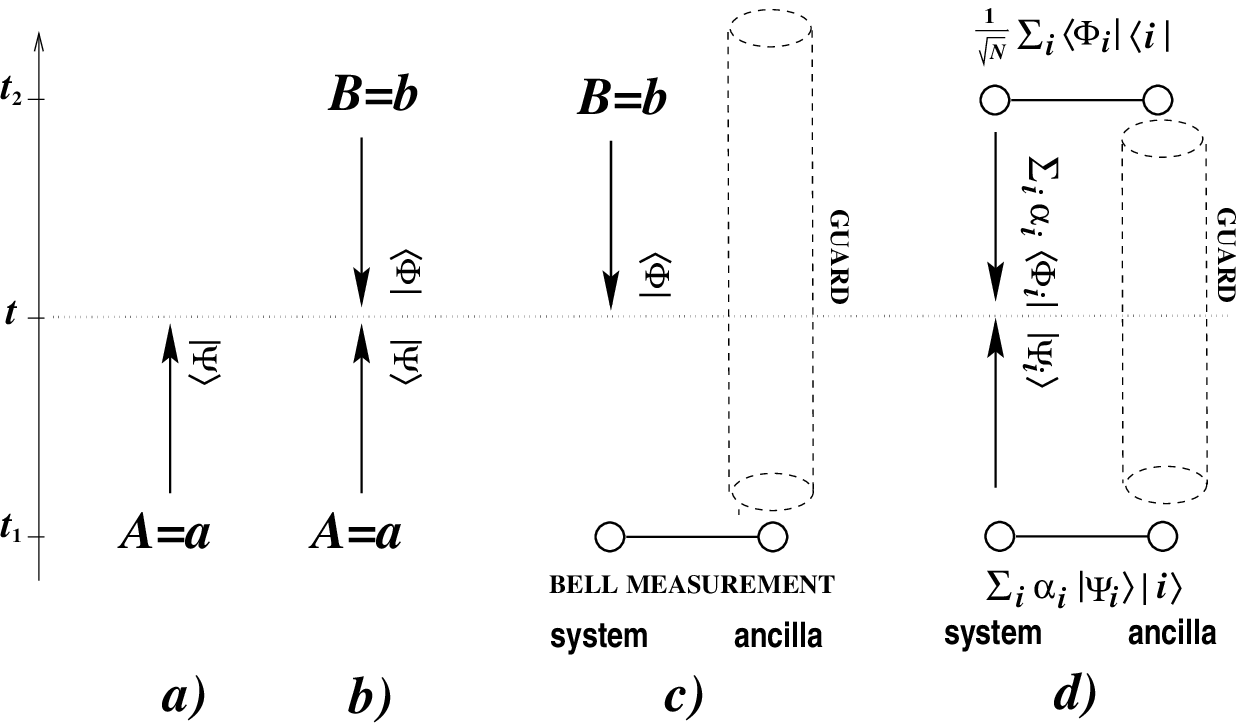}}
\vskip .2cm
\noindent
{\bf Fig. 1 Description of quantum systems:}~
 (a) pre-selected, (b) pre- and
post-selected, (c) post-selected, and (d) generalized pre- and
post-selected.
\vskip .4cm

\vskip .4cm \noindent {\bf The  two-state vector}

In order to have {\it now} a system the complete description of which
at time $t$ is a two-state vector  (\ref{2sv}), there should be a
complete measurement in the past of time $t$ and a complete
measurement after the time $t$, see  Fig. 1{\it b}.
In addition to the   measurement $A=a$ at
  time $t_1$, there should be  a complete measurement at $t_2$,
 $t_2>t$, obtaining
  a
specific outcome $B=b$  such that  the backward time  evolution
  from $t_2$
to $t$ leads to the desired  state
\begin{equation}
  \label{bra=}
\langle\Phi | ~=~\langle b|~ U^{\dagger}(t, t_2) .
\end{equation}

The time ``now'', $t_{now}$ is clearly larger than $t_2$. The two-state vector
$\langle \Phi|~|\Psi\rangle$ is the complete description of the system
at time $t$ starting
from  the time $t_2$ and forever.

\vskip .4cm \noindent {\bf A single backward-evolving quantum state}

We have presented above a description of quantum systems by
a single forward-evolving
quantum state (\ref{1qs}) and by a two-state vector  (\ref{2sv}). It is
natural to ask: Are there  systems described by a  single backward-evolving
quantum state? The notation for such a state is
\begin{equation}
\label{bra}
 \langle\Phi |.
\end{equation}
A measurement of $B$ at time $t_2$, even in the case it yields the
desired outcome $B=b$, is not enough. The difference between
preparation of (\ref{1qs}) and (\ref{2sv}) is that at present, $t$, the
future of a quantum system does not exist (the future measurements
have not been performed yet), but the past of a quantum system exists:
it seems that even if {\it we} do not know it, there is a  quantum state
of the system evolving towards the future  defined by the
results of  measurements in the past. Therefore, in order to obtain a
quantum system described by a backward evolving quantum state
(\ref{2sv}), in addition to the post-selection measurement performed
after  time $t$, we have to {\it erase} the past.

How to erase the past of  a quantum system?
A complete measurement before the time $t$ certainly partially erases
the information which the system had before the measurement, but it
also creates the new information: the result of this measurement.
It creates another quantum state evolving forward in time, and this is, really,
what we need to erase. We
have to achieve the situation in which no information  arrives from
the past. It seems impossible given the assumption that all the past
is known. However, if we  perform a measurement on a composite system
containing our system and an auxiliary system, {\it an
ancilla}, then it can be done, see Fig. 1{\it c}.
Performing a Bell-type measurement results
in one of a completely correlated states of the system and the
ancilla (the EPR-type state). In such a state, each system has equal
probability to be found in any state. However, the measurement on one
system fixes the state of the other, so, in addition to the Bell-type
measurement we need to ``guard'' the ancilla such that no measurement could
be
performed on it until now. Again, the complete description of a
quantum system by a single (this time backward-evolving) quantum
state can be achieved only for a period of time until the measurements
on the ancilla would fix the forward-evolving quantum state for the system.

The backward evolving state is a premise not only of the two-state
vector formalism, but also of ``retrodictive'' quantum mechanics
\cite{Ba1,Ba2,Ba3,retro1,retro2}, which deals with the analysis of quantum systems based
on a quantum measurement performed in the future relative to the
time in question. It is also relevant to ``consistent histories''
and ``decoherent histories'' approaches \cite{G,GH}.

\subsection{The generalized two-state vector}
\label{sec:gtsv}

The descriptions we described above correspond to an ``ideal'' case. We
have assumed that complete measurements have been performed on the system
in the past, or in the future or both. The philosophical question is this:
can we assume that going sufficiently far away to the past, far away
to the future and far in the sense of considering composite systems
larger and larger, at the end there always be a complete description
in the form
of a two-state vector. Usually we do put constraints how far we go (at least
regarding the future and the size of the system). In constructing the
situation in which a system is described by a backward evolving quantum
state only, we already limited ourselves to a particular system instead
of being satisfied by the correct claim that our system is a part of a
composite system (which includes also the ancilla) which does have
forward evolving quantum state. As in the standard approach, limiting
our analysis to a particular system leads to descriptions with {\it
mixed} states. There are situations in which the forward evolving state
is a mixed state (the system is correlated to an ancilla) and backward
evolving state is another mixed state (the system correlated to
another ancilla).  Although the generalization to the mixed states is
straightforward, it is not obvious  what is its most convenient
form. For a powerful, but somewhat cumbersome formalism, see
\cite{RA}. However, there is a particular case which is not too
difficult to describe. It corresponds to another ``pure'' two-state vector
description: {\it generalized two-state vector}.

Generalized two-state vector \cite{AV91} is the name for the
superposition of two-state vectors
\begin{equation}
  \label{g2sv}
 \sum_i  \alpha_i  \langle\Phi_i |~ |\Psi_i \rangle .
\end{equation}
In general, the sets $\{|\Psi_i\rangle \}$, $\{\langle \Phi_i |\}$
need not be orthogonal. Then, the normalization should be chosen
consistently, although it is not very important since in main
applications of this concept the normalization does not affect
anything.

For simplicity we will consider the case of zero free Hamiltonian for
the system and for the ancilla.
In order to obtain  the generalized two-state vector (\ref{g2sv}) we
have to prepare at $t_1$ the system and the ancilla in a correlated
state $ \sum_i \alpha_i |\Psi_i \rangle |i \rangle$, where $\{
|i \rangle\}$ is a set of orthonormal states of the ancilla.
Then we have to ``guard'' the ancilla such that there will be no
measurements or any
other interactions performed on the ancilla  until the post-selection
measurement of a projection on the correlated state $1/\sqrt N \sum_i~
 |\Phi_i \rangle |i \rangle $,
 see Fig. 1{\it d}. If we obtain the desired outcome, then the system is
described at time $t$ by the generalized two-state vector (\ref{g2sv}).

\section{ Ideal  Quantum Measurements}

\subsection{Von Neumann Measurements}

In this section I shall discuss how a quantum system characterized by
a certain description interacts with other systems. Some particular
types of interactions are named  {\it measurements}  and the effect of
these interactions characterized as the results of these
measurements. The basic concept is an  {\it ideal quantum measurement}
of  an observable $C$. This operation is defined for pre-selected
quantum systems in the following way:
\begin{quotation}
  If the  state of a quantum system before the measurement was
an eigenstate of $C$ with an eigenvalue $c_n$ then the outcome of the
measurement is $c_n$ and the quantum state of the system is not changed.
\end{quotation}
The standard implementation of the ideal quantum measurement
is modeled by the von Neumann
 Hamiltonian \cite{neumann}:
\begin{equation}
  \label{neumann}
 H = g(t) P C,
\end{equation}
where $P$ is the momentum conjugate to the pointer variable $Q$, and
the normalized coupling function $g(t)$ specifies the time of the
measurement interaction. The outcome of the measurement is the shift
of the pointer variable during the interaction. In an ideal
measurement the function $g(t)$ is nonzero only during a very short
period of time, and the free Hamiltonian during this period of time
can be neglected.

\subsection{The Aharonov-Bergmann-Lebowitz rule}
\label{sec:ABL}

 For a quantum system described by the two-state
vector (\ref{2sv}), the probability for an  outcome $c_n$ of an ideal
measurement of an  observable $C$ is given by
\cite{ABL,AV91}
\begin{equation}
  \label{ABL}
 {\rm Prob}(c_n) = {{|\langle \Phi | {\bf P}_{C=c_n} | \Psi \rangle |^2}
\over{\sum_j|\langle \Phi | {\bf P}_{C=c_j} | \Psi \rangle |^2}} .
\end{equation}

  For a quantum system described by a {\it generalized two-state
vector} (\ref{g2sv}) the probability for an  outcome $c_n$ is given by
\cite{AV91}
\begin{equation}
  \label{ABL-gen}
 {\rm Prob}(c_n) = {{|\sum_i \alpha_i \langle \Phi_i | {\bf P}_{C=c_n} | \Psi_i \rangle |^2}
\over{\sum_j|\sum_i \alpha_i \langle \Phi_i | {\bf P}_{C=c_j} | \Psi_i \rangle |^2}} .
\end{equation}

Another important generalization of the formula (\ref{ABL}) is for the
case in which the post-selection measurement is not complete and
therefore it does not specify a single post-selection state $\langle
\Phi|$. Such an example was recently considered by Cohen \cite{CO} in
an (unsuccessful \cite{CO-co}) attempt to
find constraints to the applicability of
the ABL formula.
In this case, the post-section measurement is a projection on a
{\em degenerate} eigenvalue of an observable $B=b$. The modified ABL formula
is \cite{CO-co}:
\begin{equation}
  \label{ABL-new}
{\rm Prob}(c_n) = {{\Vert {\bf P}_{B=b}
{\bf P}_{C=c_n} | \Psi \rangle \Vert^2}
\over{\sum_j \Vert {\bf P}_{B=b}  {\bf P}_{C=c_j} | \Psi \rangle \Vert^2}} .
\end{equation}
This form of the ABL formula allows to connect it to the standard
formalism of quantum theory in which there is no post-selection. In the
limiting case when the projection operator  ${\bf P}_{B=b}$ is just
the unity operator {\bf I}, we  obtain the usual expression:
\begin{equation}
  \label{ABL-pre}
 {\rm Prob}(c_n) = ||  {\bf P}_{C=c_n} | \Psi \rangle ||^2 .
\end{equation}

\subsection{Three-boxes example}

Consider a particle which can be located in one out of three boxes. We
denote the state of the particle when it is in box $i$ by $|i\rangle
$. At time $t_1$ the particle is prepared in the state
\begin{equation}
  \label{3psiin}
  |\Psi\rangle = {1 \over {\sqrt{3}}} (|1\rangle + |2\rangle +
|3\rangle ).
\end{equation}
At time $t_2$ the particle is found to be in the state
\begin{equation}
  \label{3psifin}
  |\Phi\rangle = {1 \over {\sqrt{3}}} (|1\rangle + |2\rangle -
|3\rangle ).
\end{equation}
We assume that in the time interval $[t_1, t_2]$ the Hamiltonian is
zero.
 Therefore, at time $t$, $t_1<t<t_2$, the particle is described by
 the two-state vector
\begin{equation}
  \label{3tsv}
 \langle \Phi|~ |\Psi\rangle = {1 \over {3}} (\langle 1|+ \langle 2| -
\langle 3|)~ (|1\rangle + |2\rangle +
|3\rangle ).
\end{equation}
Probably the most  peculiar fact about this single particle is that
it can be found with certainty in two boxes \cite{AV91}. Indeed, if at
time $t$ we open box 1, we are certain to find the particle in box 1;
and if we open box 2 instead, we are certain to find the particle in
box 2.  These results can be obtained by straightforward
application of the ABL formula (\ref{ABL}).  Opening box $i$
corresponds to measuring the projection operator
${\rm \bf P}_{i} =|i\rangle \langle
i|$.  The corresponding operators appearing in (\ref{ABL}) are
\begin{equation}
  \label{oper}
 {\bf P}_{{\bf P}_i=1} =|i\rangle \langle i|, ~~~~~~~~~ {\bf P}_{{\bf P}_i=0} = \sum_{j\neq i} |j\rangle \langle j|
\end{equation}
Therefore, the calculation of the probability to find the particle in box 1 yields:
\begin{equation}
  \label{probbox1}
 {\rm Prob}({\bf P}_1=1) = {{|\langle \Phi | 1\rangle \langle 1 | \Psi \rangle |^2}
\over{|\langle \Phi | 1\rangle \langle 1 | \Psi \rangle  |^2 + |\langle \Phi | 2\rangle \langle 2 | \Psi \rangle + \langle \Phi | 3\rangle \langle 3 | \Psi \rangle |^2 }} = {{|{1\over 3}|^2}\over{|{1\over 3}|^2 +|0|^2}} =1.
\end{equation}
Similarly, we obtain $ {\rm Prob}({\bf P}_2=1) =1$. Note, that if we open
both box 1 and box 2, we might not see the particle at all.

This example can be generalized to the case of a  large number of boxes $N$. A single particle described by a two-state vector
\begin{equation}
  \label{Ntsv}
   {1 \over {N}} (\langle 1|+ \langle 2| + ... - \sqrt{N-2}
\langle N|)~ (|1\rangle + |2\rangle + ... +\sqrt{N-2}
|N\rangle ).
\end{equation}
This  single particle is, in some sense, simultaneously in $N-1$
boxes: whatever box is opened (except the last one) we are certain
to find the particle there.

Recently, we found that the particle is simultaneously in several
boxes even in a more robust sense \cite{shut}. We cannot find it
simultaneously in all boxes if we look at all of them, but a single
photon can!  We found that a photon will scatter from our pre- and
post-selected particle, as if there were particles in all boxes.

The analysis of the three-boxes example has interesting features
also in the framework of the consistent histories approach
\cite{Grif,Kent,Grif1}. On the other hand, it generated significant
controversies. The legitimacy of counterfactual statements were
contested, see discussion in Section 5.4, the Kastner criticism
\cite{Kast} and Vaidman's reply \cite{reply}, and it was claimed by
Kirkpatrick \cite{Kirk} that the three-boxes example does not
exhibit genuine quantum paradoxical feature because it has a
classical counterpart. Very recently Ravon and Vaidman \cite{RV}
showed that Kirkpatrick's proposal fails to mimic quantum behavior
and that the three-box example is one of not too many classical
tasks which can be done better using quantum tools. (We could not
see a refutation of this statement in Kirkpatrick's reply
\cite{Kirkrep}).
 This is the
paradoxical feature of the three-box experiment which was overlooked
by Leavens et al. \cite{Leav-lev} who considered variations of the
three-box experiment with modified pre- and post-selected states.

Recently, a setup equivalent to the three-box example was presented
as a novel {\it counterfactual computation} method \cite{HON}. The
analysis of this proposal in the framework of the two-state vector
formalism \cite{CFC-3B} shows that one cannot claim that the
computer yields the result of computation without actually
performing the computation and therefore, the proposal fails to
provide counterfactual computation for all possible outcomes as it
was originally claimed.

\subsection{The failure of the product rule}

An important difference between pre- and post-selected
systems and pre-selected systems only is that the {\it product rule}
does not hold \cite{PR}. The
product rule, which does hold for pre-selected quantum systems is that
if $A=a$ and $B=b$ with certainty, then it is certain that $AB
= ab$. In the three-boxes case we know with certainty that ${\rm \bf
 P}_1=1$, ${\rm \bf P}_2=1$. However, ${\rm \bf P}_1 {\rm \bf P}_2=0$.

Another  example of this kind in a which measurement in one place affects
the outcome of a measurement in another place is a pre- and post-selected
pair of  separate spin-$1\over 2$ particles \cite{PR2}.
  The particles are  prepared, at time $t_1$, in
a singlet state.
At time $t_2$ measurements of ${\sigma_1}_x$ and ${\sigma_2}_y$ are
performed and certain results are obtained, say ${\sigma_1}_x=1$ and
${\sigma_2}_y=1$, i.e. the pair is described  at time $t$, $ t_1 < t
< t_2$, by the two-state vector
\begin{equation}
  \label{2ptsv}
{1\over {\sqrt 2}}~\langle \uparrow _x|
~\langle \uparrow _y |~(|\uparrow_z\rangle
|\downarrow_z\rangle - |\downarrow_z\rangle |\uparrow_z\rangle  ) .
\end{equation}
 If at time $t$ a measurement of ${\sigma_1}_y$ is performed (and if this is
the only measurement performed between $t_1$ and $t_2$), then the
outcome of the measurement is known with certainty: ${\sigma_1}_y(t) =
-1$.  If, instead, only a measurement of
${\sigma_2}_x$ is performed at time $t$, the result of the measurement
is also certain: ${\sigma_2}_x(t) =- 1$.  The
operators ${\sigma_1}_y$ and ${\sigma_2}_x$ obviously commute, but
nevertheless, measuring ${\sigma_2}_x(t)$ clearly disturbs the outcome
of the measurement of ${\sigma_1}_y(t)$: it is not certain anymore.

Measuring the product  ${\sigma_1}_y {\sigma_2}_x$,  is, in
principle, different from the measurement of both ${\sigma_1}_y$ and
${\sigma_2}_x$ separately.
In our example the outcome of the measurement of the product {\it
is} certain, the ABL formula (\ref{ABL}) yields ${\sigma_1}_y
{\sigma_2}_x= -1$.   Nevertheless, it does not equal the product of the results
which must come out of the measurements of
${\sigma_1}_y$ and ${\sigma_2}_x$ when every one of them is performed
without the other.

Note measurability of the product ${\sigma_1}_y
{\sigma_2}_x$ using only local interactions. Indeed,
we may write the product
as a modular sum, ${\sigma_1}_y {\sigma_2}_x = (
{\sigma_1}_y + {\sigma_2}_x){\rm mod4} - 1$. It has been shown \cite{AAV86}
 that nonlocal operators such as $({\sigma_1}_y +{\sigma_2}_x){\rm mod4}$ can be measured using
 solely local interactions.

Hardy \cite{Ha1} analyzed another very spectacular example in which an
electron and a positron are found with certainty if searched for in a
particular place, but, nevertheless, if both are searched
simultaneously, there is certainty {\it not} to find them together. Again, the
failure of the product rule explains this counterintuitive situation
and the far reaching conclusions of Hardy's paper seem not to be
warranted \cite{PR}.

The two spin-$1\over 2$ particles example with a small modification
of omitting the  measurement at time $t_2$ performed on a second
particle, but instead, ``guarding'' it  starting from time $t_1$
against any measurement, is a demonstration of obtaining a quantum
system described only  by a backward evolving quantum state
${\langle \uparrow_x|}$. The probability distribution for outcomes
of spin-component measurements performed at time $t$ is identical to
that of a particle in a pre-selected state $ |\uparrow_x \rangle$.
Note that for quantum systems which are post-selected only, the
product rule does hold.

Recently \cite{MRV} it has been shown that pre- and post-selection
allows another related peculiar feature: ``a posteriori''
realization of super-correlations maximally violating the CHSH
bound, which have been termed as Popescu-Rohrlich  boxes
\cite{PoRo}.

\subsection{Ideal measurements performed on a system described by generalized
  two-state vector}
\label{sec:imgs}

Another modification,  replacing the measurements at $t_2$ on two
particles by measurement of a nonlocal variable such as a Bell operator
on both particles and guarding the second particle between $t_1$ and
$t_2$ produces a {\it generalized two-state vector} for the
first particle. Such particles  might have a peculiar feature that the
outcome of spin component measurements is certain in a continuum of
directions. This is a surprising result because
the pre-selected particle might have definite value of spin component
at most in one direction and the particle described by two-state vector
will have definite results of spin component measurements in two
directions: one defined by pre-selection and one defined by
post-selection (the directions might coincide). For example \cite{AV91}, the
particle described by a generalized two-state vector
\begin{equation}
  \label{econe}
  \cos \chi \langle \uparrow _z|~|\uparrow_z\rangle
 -\sin \chi \langle \downarrow _z|~|\downarrow_z\rangle, ~~~~~~~~\chi
 \in \left ( 0, {\pi \over 2} \right ),
\end{equation}
will yield the outcome $\sigma_{\eta} =1$ for the cone of directions
  $\hat \eta$ making angle $\theta$ with the $z$ axes such that
  $\theta = 4 \arctan \sqrt{\tan \chi}$. This can be verified directly using the formula
(\ref{ABL-gen}), but we will bring another argument for this result below.

The generalized two-state vector is obtained when there is a
particular result of the nonlocal measurement at time $t_2$. It is
interesting that we can construct a particular measurement at time
$t_2$ such that whatever the outcome will be there will be a cone of
directions in which the spin has a definite value. These cones
intersect in general in four lines. It can be arranged that they will
``touch'' on, say $x$ axis and intersect in $y$ and $z$ axes. Then, in
all cases we will be able to ascertain the value of  $\sigma_x$,
$\sigma_y$, and   $\sigma_z$ of a single particle
\cite{s-xyz}.

The problem was also analyzed in the framework of the standard
approach \cite{Ben,Me} and after coining  the name ''The Mean King
Problem'' continued to be a topic of an extensive analysis. It has
been generalized to the spin-1 particle \cite{AE} and to a higher
dimentional case \cite{EA,A1}. The research continues until today
\cite{A2,MeKi1,MeKi2,MeKi3,MeKi4,MeKi5}. Moreover, today's
technology converted from gedanken quantum game to a real
experiment. Schulz et al. \cite{SHU} performed this experiment with
polarized photons (instead of spin $-{1\over 2}$  particles).

\section{ Weak  Measurements~~}

\subsection{Introduction}
\label{sec:Intweak}

The most interesting phenomena which can be seen in the framework of
the TSVF are related to {\it weak measurements} \cite{AV90}. A weak
measurement is a standard measuring procedure (described by the
Hamiltonian (\ref{neumann})) with weakened coupling. In an ideal
measurement the initial position of the pointer $Q$ is well localized
around zero and therefore the conjugate momentum $P$ has a very large
uncertainty which leads to a very large uncertainty in the  Hamiltonian  of the
measurement (\ref{neumann}). In a weak measurement, the initial state of
the measuring device is such that $P$ is localized around zero with
small uncertainty. This leads, of course, to a large uncertainty in $Q$ and
therefore  the measurement becomes imprecise.  However,
by performing the weak measurement on an ensemble of $N$ identical systems
we improve the precision by a factor $\sqrt N$ and in some
special cases
we can obtain  good precision even in a measurement performed on a single
system \cite{AACV}.

 The idea of weak measurements is to make the
coupling with the measuring device sufficiently weak so that the
change of the quantum state due to the measurements  can be neglected.
In fact, we require that the two-state vector is not significantly disturbed,
i.e. neither
the usual, forward  evolving quantum state, nor the  backward
evolving quantum state is changed significantly. Then, the outcome
of the measurement should be affected by both states. Indeed,
the outcome of a weak measurement of a variable $C$ performed on a
system described by the two-state vector  $\langle\Phi |~ |\Psi \rangle$
is the {\it weak value} of $C$:
\begin{equation}
C_w \equiv { \langle{\Phi} \vert C \vert\Psi\rangle
\over \langle{\Phi}\vert{\Psi}\rangle } .
\label{wv}
\end{equation}
Strictly  speaking, the readings of the pointer of the measuring device
will cluster around Re$(C_w)$. In order to find Im$(C_w)$  one should measure the shift in $P$ \cite{AV90}.

The weak value for a system described by a generalized two-state vector
(\ref{g2sv}) is \cite{AV91}:
\begin{equation}
  \label{wv-gen}
 C_w = {{\sum_i \alpha_i \langle \Phi_i | C | \Psi_i \rangle }
\over{\sum_i \alpha_i \langle \Phi_i  | \Psi_i \rangle }} .
\end{equation}

Next, let us  give  the  expression for the weak value when the
post-selection measurement is not complete. Consider a system
pre-selected
in the state  $|\Psi\rangle$ and post-selected by the
measurement of a degenerate eigenvalue $b$ of a variable  $B$. The weak value of $C$ in this case is:
\begin{equation}
  \label{wv-new}
 C_w = {{ \langle{\Psi} \vert {\bf P}_{B=b}  C \vert \Psi \rangle}
\over {\langle \Psi| {\bf P}_{B=b}\vert \Psi \rangle }} .
\end{equation}

This formula allows us to find the outcome of a weak measurement
performed on a pre-selected (only) system. Replacing ${\bf P}_{B=b}$ by
the unity operator yields the result that the weak value of a
pre-selected system in the state $|\Psi\rangle$ is the expectation
value:
\begin{equation}
  \label{wv-exp}
 C_w =  \langle{\Psi} \vert   C \vert\Psi\rangle .
\end{equation}

Let us show how the weak values emerge as the outcomes of weak
measurements. We will limit ourselves to two cases: first, the weak
value of the pre-selected state only (\ref{wv-exp}) and then, the weak
value of the system described by the two-state vector (\ref{wv}).

In the weak
measurement, as in the standard von Neumann measurement, the Hamiltonian
of the interaction with the measuring device is given by (\ref{neumann}).
The weakness of the interaction is achieved by preparing the initial
state of the measuring device in such a way that the conjugate
momentum of the pointer variable,  $P$, is small, and thus the interaction
Hamiltonian (\ref{neumann}) is small.  The initial state of the
pointer variable is  modeled by a Gaussian centered at zero:

\begin{equation}
  \label{md-in}
\Psi^{MD}_{in} (Q) =(\Delta ^2 \pi )^{-1/4} e^{ -{{Q ^2} /{2\Delta ^2}}}.
\end{equation}
The pointer is in the ``zero'' position before the measurement, i.e.
its initial probability distribution is
\begin{equation}
  \label{prob}
{\rm Prob}(Q) = (\Delta ^2 \pi )^{-1/2} e^{ -{{Q ^2} /{\Delta ^2}}}.
\end{equation}

If the initial state of the system is a superposition
$
|\Psi \rangle = \Sigma \alpha_i |c_i \rangle
$,
then after the interaction (\ref{neumann}) the state of the system and the measuring device is:
\begin{equation}
  \label{st-symd}
(\Delta ^2 \pi )^{-1/4} \Sigma \alpha_i |c_i \rangle e^{ -{{(Q-c_i)^2} /{2\Delta ^2}}}.
\end{equation}
The probability distribution of the pointer variable corresponding to the
state (\ref{st-symd}) is:
\begin{equation}
  \label{prob2}
{\rm Prob}(Q) =(\Delta ^2 \pi )^{-1/2} \Sigma |\alpha_i|^2
 e^{ -{{(Q-c_i) ^2} /{\Delta ^2}}}.
\end{equation}
In case of the ideal measurement, this is a weighted sum of the initial
probability distribution localized around various eigenvalues.
Therefore,  the reading of the
pointer variable in the end of the measurement almost always yields  the
value close to one of the eigenvalues.
The limit of weak measurement corresponds to  $\Delta \gg c_i$ for all  eigenvalues
$c_i$. Then,  we can  perform the Taylor expansion of the sum (\ref{prob2}) around
$Q=0$ up to
the first order and rewrite the probability distribution of the
pointer in the following way:
\begin{eqnarray}
  \label{prob22}
 {\rm Prob}(Q) =(\Delta ^2 \pi )^{-1/2} \Sigma |\alpha_i|^2 e^{ -{{(Q-c_i)^2}
/{\Delta ^2}}} =~~~~~~~~~~~~~~~~~~~~~~~~~~~~~~~~~\nonumber\\
(\Delta ^2 \pi )^{-1/2} \Sigma |\alpha_i|^2 (1
-{{(Q-c_i) ^2} /{\Delta ^2}}) = (\Delta ^2 \pi )^{-1/2} e^{
 -{{(Q-\Sigma|\alpha_i|^2c_i) ^2} /{\Delta ^2}}} .
\end{eqnarray}
But this is exactly the initial distribution shifted by the value
$\Sigma|\alpha_i|^2c_i$.
 This is the outcome of the measurement, in this case the weak value
 is the expectation value:
 \begin{equation}
   \label{wv=ev}
   C_w =
\Sigma|\alpha_i|^2c_i =\langle \Psi |C|\Psi\rangle .
 \end{equation}
 The weak value is
obtained from statistical analysis of the readings of the measuring
devices of the measurements on an ensemble of identical quantum
systems. But it is different conceptually from the standard definition
of expectation value which is a mathematical concept defined from the
statistical analysis of the {\em ideal} measurements of the variable
$C$ all of which yield one of the eigenvalues $c_i$.

Now let us turn to the system described by the two-state vector (\ref{2sv}).
As
usual, the free Hamiltonian is assumed to be zero so it can be obtained
by  pre-selection of $|\Psi \rangle$ at $t_1$ and post-selection of
$|\Phi \rangle$ at $t_2$. The  (weak)
measurement interaction of the form (\ref{neumann}) takes place at
time $t$, $t_1<t<t_2$.   The
state of the measuring device after this sequence
of measurements is given (up to
normalization) by
\begin{equation}
  \label{st2sv}
  \Psi^{MD} (Q) = \langle \Phi \vert
e^{-iPC}
\vert \Psi \rangle e^{ -{{Q ^2} /{2\Delta ^2}}}.
\end{equation}
After simple algebraic manipulations we can rewrite it (in the
$P$-representation) as
\begin{eqnarray}
  \label{st2sv1}
\tilde  \Psi^{MD} (P) &=&   \langle \Phi
\vert
 \Psi \rangle ~ e^{-i {C_w} P} ~
e^{-{{\Delta}^2 {P^2}} /{2}} \\
\nonumber
\hfill{} & &+ \langle \Phi \vert
 \Psi \rangle  \sum_{n=2}^\infty {{(iP)^n}\over{n!}}
[(C^n)_w - (C_w)^n]   e^{ -{{\Delta ^2 P^2}} /{2}}.
\end{eqnarray}
\noindent
If $\Delta$ is sufficiently large,  we
can neglect the second term of (\ref{st2sv1}) when we Fourier transform
 back to the  $Q$-representation.  Large $\Delta$
corresponds to weak measurement in the sense that the
interaction Hamiltonian
(\ref{neumann}) is small.  Thus, in the limit of weak measurement, the final state
of the measuring device (in the $Q$-representation) is
\begin{equation}
  \label{stfin}
\Psi^{MD} (Q) = (\Delta^2 \pi )^{-1/4} e^{ -{{(Q - C_w)^2} /{2\Delta
^2}}}.
\end{equation}
This state represents a measuring device pointing to the weak value (\ref{wv}).

 Weak measurements on pre- and post-selected ensembles yield, instead
 of eigenvalues, a value which might lie far outside the range of the
 eigenvalues.  Although we have
 shown this result for a specific von Neumann model of measurements,
 the result is completely general: any coupling of a pre- and
 post-selected system to a variable $C$, provided the coupling is
 sufficiently weak, results in effective coupling to $C_w$. This weak
 coupling between a single system and the measuring device will not,
 in most cases, lead to a distinguishable shift of the pointer
 variable, but collecting the results of measurements on an ensemble
 of pre- and post-selected systems will yield the weak values of a
 measured variable to any desired precision.

When the strength of the coupling to the measuring device goes to
zero, the outcomes of the measurement invariably yield the weak
value. To be more precise, a measurement yields the real part of the
weak value. Indeed, the weak value is, in general, a complex number,
but its imaginary part will contribute only a (position dependent)
phase to the wave function of the measuring device in the position
representation of the pointer. Therefore, the imaginary part will not
affect the probability distribution of the pointer position which is
what we see in a usual measurement.  However, the imaginary part of
the weak value also has physical meaning. It is equal to the shift of
the Gaussian wave
function of the measuring device  in the  momentum
representation. Thus, measuring the shift of the momentum of the pointer will
yield the imaginary part of the weak value.

The research of weak measurements continues until today. Recently,
Botero \cite{Bot} noted that in some cases the pointer of the weak
measurements in some cases has narrower distribution after the weak
measurement interaction than it has before. Note also recent
different ways of the analysis of weak measurement effect
\cite{Abot,ta,park,Joh1,Joh2,Joh3,brun}.

\subsection{ Examples: Measurements of  spin components.}

Let us consider a simple Stern-Gerlach experiment: measurement of a
spin component of a spin-$1\over 2$ particle. We shall consider a particle
prepared in the initial state spin ``up'' in the $\hat{x}$ direction
and post-selected to be ``up'' in the $\hat{y}$ direction. At the
intermediate time we measure, weakly, the spin component in the
$\hat{\xi}$ direction which is bisector of $\hat{x}$ and $\hat{y}$,
i.e., $ \sigma_\xi = (\sigma_x + \sigma_y)/\sqrt 2 $. Thus ${|}\Psi
\rangle =|{\uparrow_x} \rangle$, $|\Phi \rangle =|{\uparrow_y}
\rangle$, and the weak value of $\sigma_\xi$ in this case is:
\begin{equation}
  \label{wsx}
  (\sigma_\xi)_w = {{\langle{\uparrow_y} |\sigma_\xi |{\uparrow_x}
\rangle} \over {\langle{\uparrow_y} |{\uparrow_x} \rangle}} =
{1\over\sqrt 2}{{\langle{\uparrow_y} | (\sigma_x + \sigma_y)
|{\uparrow_x} \rangle} \over {\langle{\uparrow_y} |{\uparrow_x}
\rangle}} = \sqrt 2 .
\end{equation}
 This value is, of course,
``forbidden'' in the standard interpretation where a spin component
can obtain the (eigen)values $\pm1$ only.

An effective Hamiltonian for measuring $\sigma_\xi$ is
\begin{equation}
  \label{effH}
H = g(t) P \sigma_\xi .
\end{equation}
  Writing the initial state of the particle in the $\sigma_\xi$
representation, and assuming the initial state (\ref{md-in}) for the measuring
device, we obtain that after the measuring interaction the quantum
state of the system and the pointer of the measuring device is
 \begin{equation}
  \label{spomd}
 \cos {(\pi/8)} |{\uparrow_\xi} \rangle e^{ -{{(Q-1)^2} /{2\Delta ^2}}} + i \sin
{(\pi/8)} |{\downarrow_\xi} \rangle  e^{ -{{(Q+1)^2} /{2\Delta ^2}}}.
\end{equation}

\vskip .5cm
\epsfxsize=9cm
 \centerline{\epsfbox{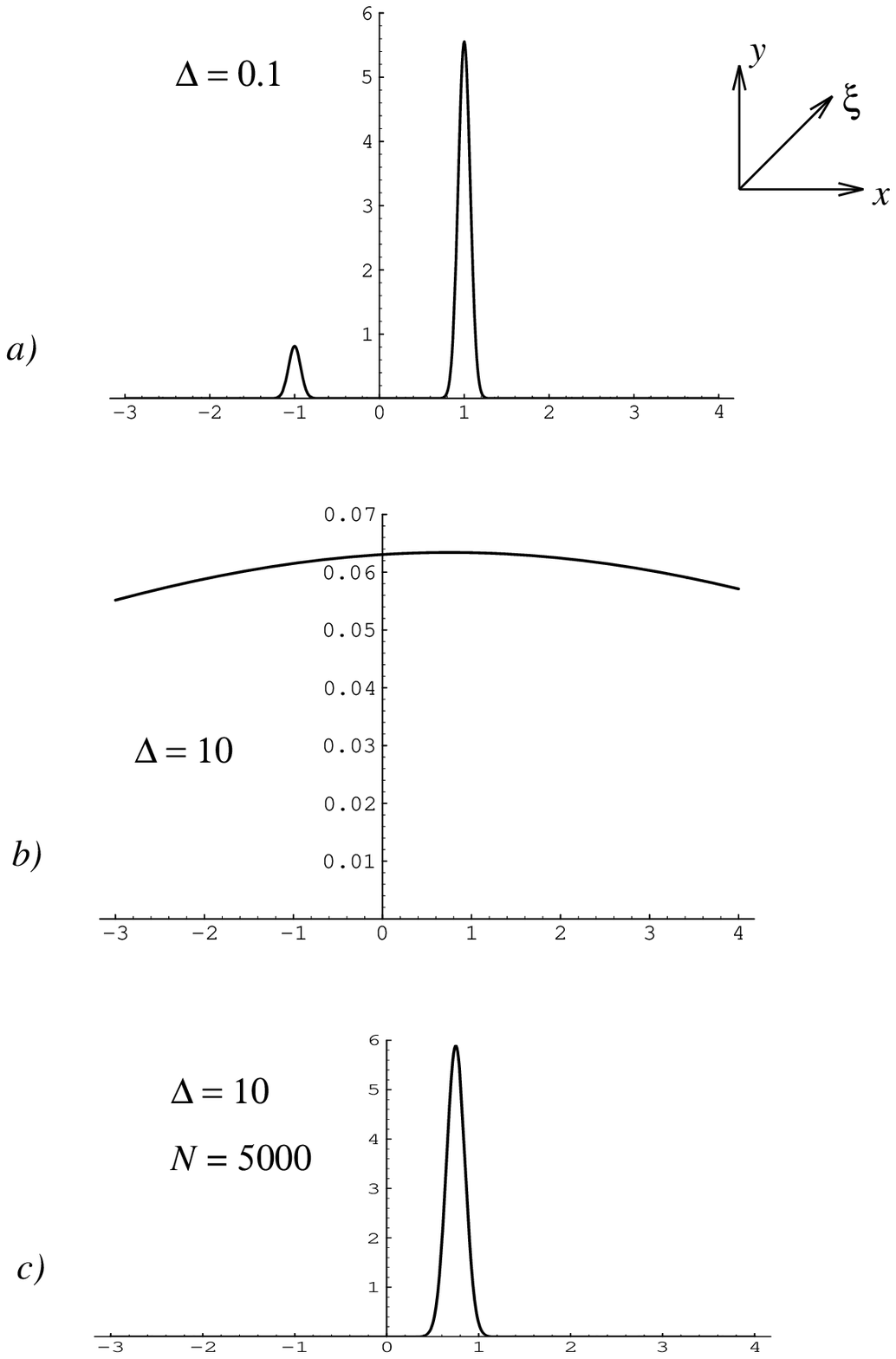}}
\vskip .2cm
\noindent
{\small {\bf
Fig. 2.~ Spin component measurement without post-selection.}~ Probability distribution of the pointer variable for measurement of
$\sigma_\xi$ when the particle is pre-selected in the state $\vert
{\uparrow_x} \rangle$.  ($a$) Strong measurement, $\Delta = 0.1.$
($b$) Weak  measurement, $\Delta = 10$.  ($c$) Weak measurement on the ensemble
of 5000 particles. The original width of the peak, 10, is reduced to
$10/\sqrt 5000 \simeq 0.14$. In the strong measurement ($a$) the pointer is localized
around the eigenvalues $\pm1$, while in the weak measurements ($b$) and ($c$)
the peak is located  in the expectation value $\langle {\uparrow_x}|
\sigma_\xi|{\uparrow_x} \rangle = 1/\sqrt2$.}

\vskip .4cm

\noindent
 The probability  distribution of the pointer position, if it is observed now
without post-selection,  is the sum of the distributions
for each spin value. It is, up to normalization,
\begin{equation}
  \label{prodis}
 {\rm Prob}(Q) = \cos^2{(\pi/8)} e^{ -{{(Q-1)^2} /{\Delta ^2}}} + \sin^2
{(\pi/8)}  e^{ -{{(Q+1)^2} /{\Delta ^2}}}.
\end{equation}
 In the usual strong measurement, $\Delta \ll 1$. In this case, as
shown on Fig. $2a$, the probability distribution of the pointer is
localized around $-1$ and $+1$ and it is strongly correlated to the
values of the spin, $\sigma_z = \pm1$.


Weak measurements correspond to a $\Delta$ which is much larger than
the range of the eigenvalues, i.e., $\Delta \gg 1$. Fig. $2b$ shows
that the pointer distribution has a large uncertainty, but it is
peaked between the eigenvalues, more precisely, at the expectation
value $\langle{\uparrow_x} |\sigma_\xi |{\uparrow_x} \rangle = 1/\sqrt
2$.  An outcome of an individual measurement usually will not be close
to this number, but it can be found from an ensemble of such
measurements, see Fig. $2c$. Note, that we have not yet considered the
post-selection.

In order to simplify the analysis of measurements on the pre- and
post-selected ensemble, let us assume that we first make the
post-selection of the spin of the particle and only then look at the
pointer of the device that weakly measures $\sigma_\xi$. We must get
the same result as if we first look at the outcome of the weak
measurement, make the post-selection, and discard all readings of
the weak measurement corresponding to the cases in which the result
is not $\sigma_y =1$. The post-selected state of the particle in the
$\sigma_\xi$ representation is $ {\langle{\uparrow_y}| = \cos
{(\pi/8)} \langle{\uparrow_\xi} | - i\sin {(\pi/8)}
\langle{\downarrow_\xi} | }$.  The state of the measuring device
after the post-selection of the spin state is obtained by projection
of (\ref{spomd}) onto the post-selected  spin state:
 \begin{equation}
  \label{psss}
 \Phi (Q) ={\cal N} \Bigl(\cos^2 {(\pi/8)} e^{ -{{(Q-1)^2}
/{2\Delta ^2}}} - \sin^2 {(\pi/8)} e^{ -{{(Q+1)^2} /{2\Delta
^2}}})\Bigr),
\end{equation}
 where ${\cal N}$ is a normalization factor. The probability
distribution of the pointer variable is given by
\begin{equation}
  \label{prdis1}
{\rm Prob}(Q) ={\cal N}^2 \Bigl (\cos^2 {(\pi/8)} e^{ -{{(Q-1)^2}
/{2\Delta ^2}}} - \sin^2 {(\pi/8)} e^{ -{{(Q+1)^2} /{2\Delta
^2}}})\Bigr)^2 .
\end{equation}

 If the measuring interaction is strong, $\Delta \ll 1$, then the
distribution is localized around the eigenvalues $\pm 1$ (mostly
around 1 since the pre- and post-selected probability to find
$\sigma_\xi =1$ is more than 85\%), see Figs. $3a,3b$. But when the
strength of the coupling is weakened, i.e., $\Delta$ is increased, the
distribution gradually changes to a single broad peak around $\sqrt
2$, the weak value, see Figs. $3c-3e$.

The width of the peak is large and therefore each individual reading
of the pointer usually will be far from $\sqrt 2$. The physical
meaning of the weak value can, in this case,  be associated only with
an ensemble of pre- and post-selected particles. The accuracy of
defining the center of the distribution goes as $1/\sqrt N$, so
increasing $N$, the number of particles in the ensemble, we can find
the weak value with any desired precision, see Fig. $3f$.

In our example, the weak value of the spin component is $\sqrt 2$,
which is only slightly more than the maximal eigenvalue, 1. By
appropriate choice of the pre- and post-selected states we can get
pre- and post-selected ensembles with arbitrarily large weak value of
a spin component. One of our first proposals [6] was to obtain
$(\sigma_\xi)_w = 100$. In this case the post-selected state is nearly
orthogonal to the pre-selected state and, therefore, the probability
to obtain appropriate\break
\vskip .5cm
\epsfxsize=12cm
 \centerline{\epsfbox{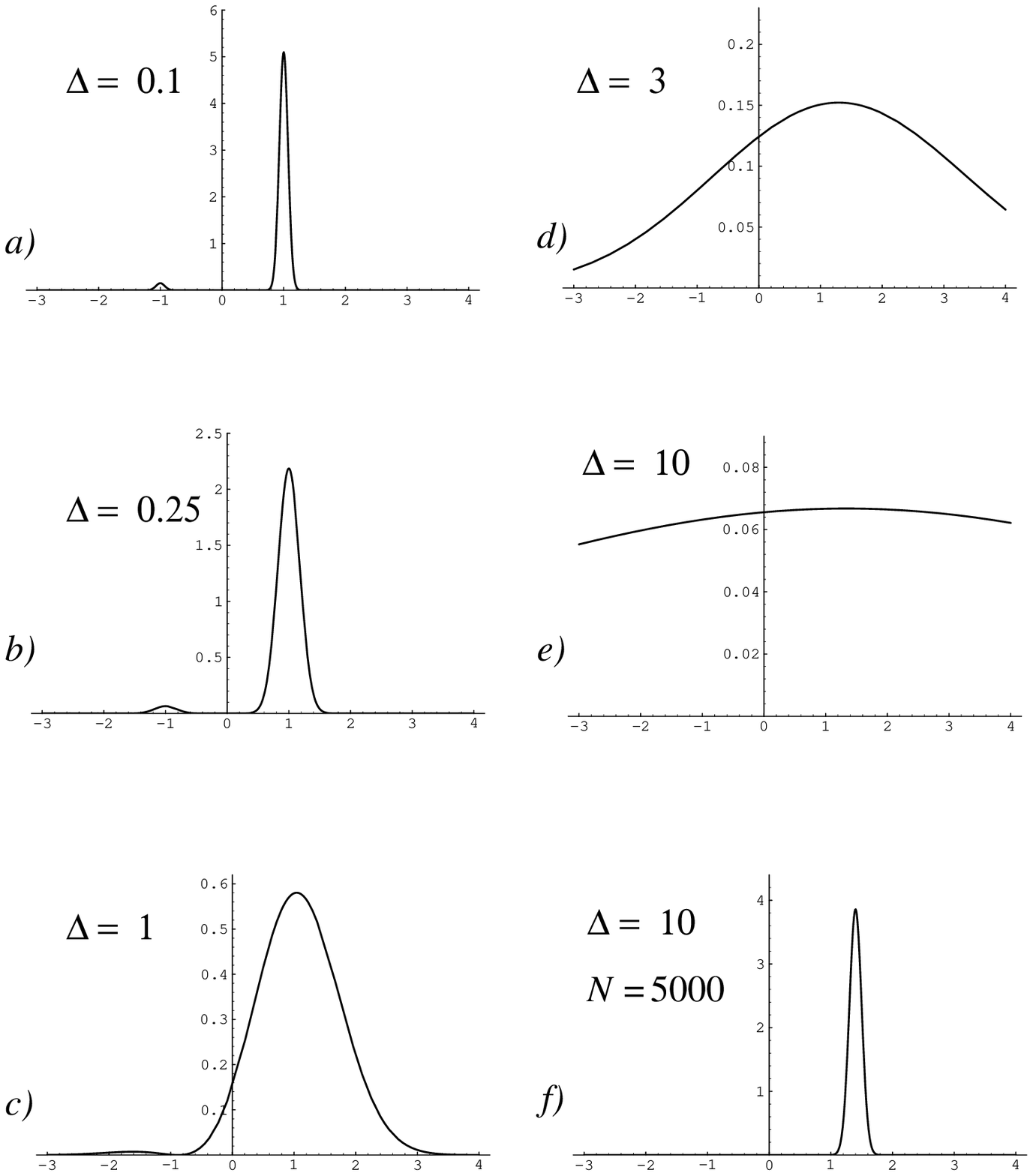}}
\vskip .2cm
\noindent
{\small {\bf
Fig. 3.~ Measurement on pre- and post-selected ensemble.}~
 Probability distribution of the pointer variable for measurement of
$\sigma_\xi$ when the particle is pre-selected in the state $\vert
{\uparrow_x} \rangle$ and post-selected in  the state $\vert
{\uparrow_y} \rangle$. The  strength of the
measurement  is
parameterized by the width of the distribution $\Delta$.
 ($a$) $\Delta = 0.1$; ($b$) $\Delta = 0.25$; ($c$) $\Delta =
1$; ($d$) $\Delta = 3$; ($e$) $\Delta = 10$.
  ($f$) Weak measurement on the ensemble
of 5000 particles; the original width of the peak, $\Delta = 10$, is reduced to
$10/\sqrt 5000 \simeq 0.14$. In the strong measurements ($a$)-($b$)
the pointer is localized
around the eigenvalues $\pm1$, while in the weak measurements ($d$)-($f$)
the peak of the distribution is located in the weak value
$(\sigma_\xi)_w = \langle {\uparrow_y}|
\sigma_\xi |{\uparrow_x} \rangle/\langle {\uparrow_y}|{\uparrow_x} \rangle
= \sqrt2$. The outcomes of the weak measurement on the ensemble
of 5000 pre- and post-selected particles, ($f$), are clearly outside the
range of the eigenvalues, (-1,1).}

\vskip .5cm
\epsfxsize=10cm
 \centerline{\epsfbox{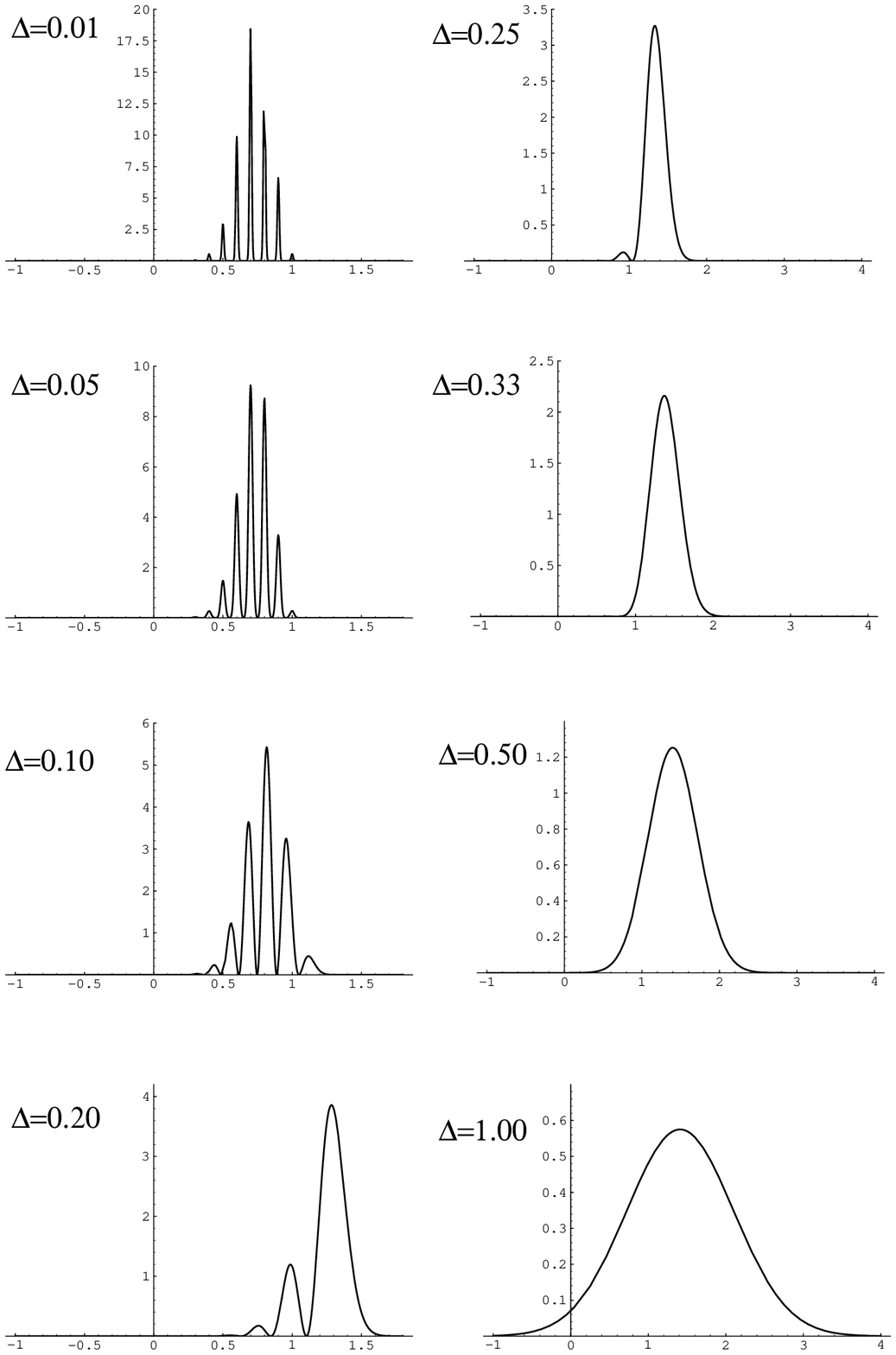}}
\vskip .2cm
\noindent
{\small{\bf Fig. 4.~ Measurement on a single system.}~ Probability
  distribution of the pointer variable for the measurement of $A
  =(\sum_{i=1}^{20} (\sigma_i)_\xi)/20$ when the system of 20 spin-$1\over 2$
  particles is pre-selected in the state $|\Psi_1\rangle =
  \prod_{i=1}^{20} |{\uparrow_x} \rangle _i$ and post-selected in the
  state ${|\Psi_2\rangle = \prod_{i=1}^{20} |{\uparrow_y} \rangle _i}$.
  While in the very strong measurements, $\Delta = 0.01-0.05$, the
  peaks of the distribution located at the eigenvalues, starting from
  $\Delta = 0.25$ there is essentially a single peak at the location
  of the weak value, $A_w = \sqrt 2$.}

\vskip .4cm

\noindent
post-selection becomes very small. While in the
case of $(\sigma_\xi)_w = \sqrt 2$ the pre- and post-selected
ensemble was about half of the pre-selected ensemble, in the case of
$(\sigma_\xi)_w = 100$ the post-selected ensemble will be smaller than
the original ensemble by the factor of $\sim 10^{-4}$.

\subsection{Weak measurements which are not really weak.}

We have shown that weak measurements can yield very surprising values
which are far from the range of the eigenvalues. However, the
uncertainty of a single weak measurement (i.e., performed on a single
system) in the above example is larger than the deviation from the
range of the eigenvalues. Each single measurement separately yields
almost no information and the weak value arises only from the
statistical average on the ensemble. The weakness and the uncertainty of
the measurement goes together. Weak measurement corresponds to small
value of $P$ in the Hamiltonian (\ref{neumann}) and,
therefore, the uncertainty in
$P$ has to be small.  This requires large $\Delta$, the uncertainty of
the pointer variable. Of course, we can construct measurement with
large uncertainty which is not weak at all, for example, by preparing
the measuring device in a mixed state instead of a Gaussian, but no
precise measurement with weak coupling is possible. So, usually, a
weak measurement on a single system will not yield the weak value
with a good precision. However, there are special cases when it is not
so. Usual strength measurement on a single pre- and post-selected
system can yield ``unusual'' (very different from the eigenvalues)
weak value with a good precision. Good precision means that the
uncertainty is much smaller than the deviation from the range of the
eigenvalues.

Our example above was not such a case. The weak value $(\sigma_\xi)_w
= \sqrt 2$ is larger than the highest eigenvalue, 1, only by $\sim
0.4$, while the uncertainty, 1, is not sufficiently large for
obtaining the peak of the distribution near the weak value, see
Fig. 3$c$.  Let us modify our experiment in such a way that a single
experiment will yield meaningful surprising result.  We consider a
system of $N$ spin-$1\over 2$ particles all prepared in the state
$|{\uparrow_x} \rangle$ and post-selected in the state $|{\uparrow_y}
\rangle$, i.e., $|\Psi\rangle = \prod_{i=1}^N |{\uparrow_x} \rangle
_i$ and $\langle \Phi | = \prod_{i=1}^N \langle{\uparrow_y}|_i$.
The variable which is measured at the intermediate time is $C
\equiv(\sum_{i=1}^N (\sigma_i)_\xi)/N$. The operator $C$ has $N+1$
eigenvalues equally spaced between $-1$ and $+1$, but the weak value
of $C$ is
\begin{equation}
  \label{wvC}
  C_w = {{\prod_{k=1}^N \langle{\uparrow_y}|_k ~ \sum_{i=1}^N
((\sigma_i)_x + (\sigma_i)_y) ~ \prod_{j=1}^N |{\uparrow_x} \rangle_j}
\over { \sqrt 2 ~ N(\langle{\uparrow_y} |{\uparrow_x} \rangle)^N}} =
\sqrt 2 .
\end{equation}
The interaction Hamiltonian is
\begin{equation}
  \label{Hint}
H = {{g(t)}\over N} P \sum_{i=1}^N  (\sigma_i)_\xi.
\end{equation}
 The initial state of the measuring device defines the precision of
the measurement. When we take it to be the Gaussian (6), it is
characterized by the width $\Delta$. For a meaningful experiment we
have to take $\Delta$ small. Small $\Delta$ corresponds to large
uncertain $P$, but now, the strength of the coupling to each
individual spin is reduced by the factor $1/N$. Therefore, for large
$N$, both the forward-evolving state and the backward-evolving state
are essentially not changed by the coupling to the measuring
device. Thus, this single measurement yields the weak value.  In
Ref. 7 it is proven that if the measured observable is an average on a
large set of systems, $C = \bigl(\sum_i^N C_i \bigr)/N$, then we can
always construct a single, good-precision measurement of the weak
value.  Here let us present just numerical calculations of the
probability distribution of the measuring device for $N$ pre- and
post-selected spin-$1\over 2$ particles. The state of the pointer after the
post-selection for this case is

\begin{equation}
  \label{stfin1}
{\cal N}\sum_{i=0}^N {(-1)^i \over (i!(N-i)!)} \bigl(\cos ^2
(\pi/8)\bigr)^{N-i} ~\bigl(\sin ^2 (\pi/8)\bigr)^i
~e^{-(Q-{{(2N-i)}\over N})^2/{2\Delta^2}} .
\end{equation}
The probability distribution for the pointer variable $Q$ is
\begin{equation}
  \label{probfin}
 prob (Q) ={\cal N}^2 \Bigl (\sum_{i=0}^N  {(-1)^i \over (i!(N-i)!)}\bigl(\cos ^2
(\pi/8)\bigr)^{N-i} \bigl(\sin ^2 (\pi/8)\bigr)^i
e^{-(Q-{{(2N-i)}\over N})^2/{2\Delta^2}}\Bigr)^2 .
\end{equation}
 The results for $N=20$ and different values of $\Delta$ are
presented in Fig. 4. We see that for $\Delta = 0.25$ and larger, the
obtained results are very good: the final probability distribution of
the pointer is peaked at the weak value, $\bigl((\sum_{i=1}^N
(\sigma_i)_\xi)/N\bigr)_w = \sqrt 2$. This distribution is very close
to that of a measuring device measuring operator $O$ on a system in an
eigenstate $|O{=} \sqrt2 \rangle$.  For $N$ large, the relative
uncertainty can be decreased almost by a factor $1/\sqrt N$ without
changing the fact that the peak of the distribution points to the weak
value.

Although our set of particles pre-selected in one state and
post-selected in another state is considered as one system, it looks
 like an ensemble.  In quantum theory, measurement of the sum
does not necessarily yield the same result as the sum of the results
of the separate measurements, so conceptually our measurement on the
set of particles differs from the measurement on an ensemble of pre-
and post-selected particles. However, in our example of weak
measurements, the results are the same.

A less ambiguous case is the example considered in the first work on
weak measurements \cite{AACV}. In this work a single system of a large spin
$N$ is considered. The system is pre-selected in the state
$|\Psi\rangle = |S_x {=} N \rangle$ and post-selected in the state
$|\Phi\rangle = |S_y {=} N \rangle$.  At an intermediate time the
spin component $S_\xi$ is weakly measured and again the ``forbidden"
value $\sqrt 2 N$ is obtained. The uncertainty has to be only slightly
larger than $\sqrt N$.  The probability distribution of the results is
centered around $\sqrt 2 N$, and for large $N$ it lies clearly outside
the range of the eigenvalues, $(-N, N)$. Unruh \cite{Unruh} made computer
calculations of the distribution of the pointer variable for this case
and got results which are very similar to what is presented on Fig. 4.

An even more dramatic example is a measurement of the kinetic energy
of a tunneling particle \cite{K<0}. We consider a particle pre-selected in a
bound state of a potential well which has negative potential near the
origin and vanishing potential far from the origin; $|\Psi\rangle =
|E{=} E_0 \rangle$.  Shortly later, the particle is post-selected to
be far from the well, inside a classically forbidden tunneling region;
this state can be characterized by vanishing potential $|\Phi\rangle
= |U {=} 0 \rangle$. At an intermediate time a measurement of the
kinetic energy is performed. The weak value of the kinetic energy in
this case is
\begin{equation}
  \label{K<0}
  K_w =  {{\langle U{=}0 | K | E {=} E_0 \rangle}
\over{\langle U{=}0 | E{=}E_0 \rangle} } =
  {{\langle U{=}0 | E - U | E {=} E_0 \rangle}
\over{\langle U{=}0 | E{=}E_0 \rangle} } = E_0 .
\end{equation}
 The energy of the bound state, $E_0$, is negative, so the weak
value of the kinetic energy is negative.  In order to obtain this
negative value the coupling to the measuring device need not be too
weak. In fact, for any finite strength of the measurement we can
choose the post-selected state sufficiently far from the well to
ensure the negative value. Therefore, for appropriate post-selection,
the usual {\bf \it strong} measurement of a positive definite
operator invariably yields a negative result!  { This weak value
predicted by the two-state vector formalism demonstrates a remarkable
consistency: the value obtained is exactly the value that we would
expect a particle to have when the particle is characterized in the
intermediate times by the two wave-functions, one in a ground state, and the
other localized outside the well.  Indeed, we obtain this result
precisely when we post-select the particle far enough from the well
that it could not have been kicked there as a result of the
intermediate measurement.   A peculiar interference
 effect of the pointer takes place: destructive interference in the
 whole ``allowed" region and constructive interference of the tails in
 the ``forbidden" negative region.  The initial state of the measuring
 device $\Phi (Q)$, due to the measuring interaction and the
 post-selection, transforms into a superposition of shifted wave
 functions. The shifts are by the (possibly small) eigenvalues, but
 the superposition is approximately equal to the original wave
 function shifted by a (large and/or forbidden) weak value:
\begin{equation}
\label{sup1}
 \sum_n  \alpha_n \Psi^{MD} (Q - c_n) \simeq  \Psi^{MD}  (Q - C_w) .
\end{equation}

These surprising, even paradoxical effects are really only gedanken
experiments. The reason is that, unlike weak measurements on an
ensemble, these are extremely rare events. For yielding an unusual
weak value, a single pre-selected system needs an extremely improbable
outcome of the post-selection measurement.  Let us compare this with a
weak measurement on an ensemble. In order to get $N$ particles in a
pre- and post-selected ensemble which yield $(\sigma_\xi)_w = 100$, we
need $\sim N 10^4$ particles in the pre-selected ensemble. But, in
order to get a single system of $N$ particles yielding $(S_\xi)_w =
100 N$, we need $\sim 10^{4N}$ systems of $N$ pre-selected
particles. In fact, the probability to obtain an unusual value by
error is much larger than the probability to obtain the proper
post-selected state.  What makes these rare effects interesting is
that there is a strong (although only one-way) correlation: for
example, every time we find in the post-selection measurement the
particle sufficiently far from the well, we know that the result of
the kinetic energy is negative, and not just negative: it is equal to
the weak value, $K_w = E_0$, with a good precision.

\subsection{Relations between weak and strong measurements}
\label{sec:relation}

In general, weak and strong measurements do not yield the same
outcomes. The outcomes of strong measurements are always the
eigenvalues while the outcomes of weak measurements, the weak values,
might be very different from the eigenvalues. However, there are two
important relations between them \cite{AV91}.

{\bf (i)} If the description of a quantum system is such that a particular
eigenvalue of a variable is obtained with certainty in case it
is measured strongly, then the weak value of this variable is equal to
this eigenvalue. This is correct in all cases, i.e., if the system
described by a corresponding single  (forward or backward evolving)
eigenstate, or if it is described by a two-state vector, or even if it
is described by a generalized two-state vector.

{\bf (ii)} The inverse of this theorem is true for dichotomic
variables such as projection operators of spin components of spin-$1\over 2$
particles. The proofs of both statements are given in \cite{AV91}.

Let us apply the theorem (i) for the example of 3 boxes when we have a
large number of particles all pre- and post-selected in the two-state
vector (\ref{3tsv}).
The {\em actual} story is as follows:
 A macroscopic number $N$ of particles (gas) were all prepared at
$t_1$ in a superposition of being in three separated boxes
(\ref{3psiin}).
 At later time $t_2$ all the particles were found in another
superposition (\ref{3psifin}) (this is an extremely rare event).
In between, at time $t$, {\it weak measurements} of a number of
particles in each box, which are, essentially, usual measurements of
pressure in each box, have been performed.
 The readings of the
measuring devices for the pressure in the boxes $1$, $2$ and $3$ were
\begin{eqnarray}
\label{p-p}
\nonumber
p_1 = p,\\
p_2 = p,\\
\nonumber
p_3 = -p,
\end{eqnarray}
\noindent
where $p$ is the pressure which is expected to be in a box with $N$
particles.

We are pretty certain that this ``actual'' story never took place
because the probability for the successful post-selection  is of the
order of $3^{-N}$; for a macroscopic number $N$ it is too small for
any real chance to see it happening. However, given that the
post-selection  does happen, we are safe to claim that
the results (\ref{p-p}) are
correct, i.e., the measurements of pressure at the intermediate time
with very high probability  have shown these results.

Indeed,
the system of all particles at time $t$ (signified by index $i$) is
described by the two-state vector
\begin{equation}
\langle \Phi|~|\Psi\rangle = {{1\over
  3^N}}\prod_{i=1}^{i=N}
(\langle 1|_i + \langle 2|_i - \langle 3|_i)~
 \prod_{i=1}^{i=N}(|1\rangle_i + |2\rangle_i + |3\rangle_i) .
\end{equation}
Then, intermediate measurements yield, for each particle, probability 1 for
the the following outcomes of measurements:
\begin{eqnarray}
\label{CFS}
\nonumber
{\rm \bf P}_1 = 1 ,\\
{\rm \bf P}_2 = 1 ,\\
\nonumber{\rm \bf P}_1 + {\rm \bf P}_2 + {\rm \bf P}_3 = 1 ,
\end{eqnarray}
where ${\rm \bf P}_1$ is the projection operator on the state of the
particle in box $1$, etc.
 Thus,
from (\ref{CFS}) and theorem (i) it follows:
\begin{eqnarray}
\label{wv'}
\nonumber({\rm \bf P}_1)_w = 1 ,\\
({\rm \bf P}_2)_w = 1 ,\\
\nonumber({\rm \bf P}_1 + {\rm \bf P}_2 + {\rm \bf P}_3)_w = 1 .
\end{eqnarray}
Since  for any variables,  $(X+Y)_w = X_w +Y_w$  we can  deduce that
$({\rm \bf P}_3)_w = -1$.

 Similarly,  for the   ``number operators'' such as
${\cal N}_1 \equiv \Sigma_{i=1}^{N}{\rm \bf P}_1^{(i)}$,
where ${\rm \bf P}_1^{(i)}$ is the projection operator on the box $1$
for a particle $i$, we obtain:
\begin{eqnarray}
\nonumber({\cal N}_1)_w = N ,\\
({\cal N}_2)_w = N ,\\
\nonumber({\cal N}_3)_w = -N .
\end{eqnarray}

In this rare situation  the ``weak measurement'' need not be very weak:
a usual measurement of pressure is a weak
measurement of the number operator. Thus, the time-symmetrized
formalism yields  the surprising result (\ref{p-p}):
the result of the pressure measurement in
box $3$ is negative! It equals minus the pressure measured in
the  boxes $1$ and $2$.

Of course, the negative pressure was not measured in a real
laboratory (it requires an extremely improbable post-selection), but
a non-robust weak measurement for three-box experiment has been
performed in a laboratory \cite{3box-ex}.

 Another example of relation between strong and weak
measurements is Hardy's paradox \cite{Ha1}. The analysis of strong
measurements appears in \cite{PR} and the weak measurements are
analyzed in detail in \cite{Ha-Aetal}. See also discussions of a
realistic experimental proposals \cite{Mo,Ha-ex1,Ha-ex2,Ha-ex3}.

An application of the inverse theorem yields an alternative proof of the
results regarding strong measurements of spin components of a
spin-$1\over 2$ particle described by the generalized two-state vector
(\ref{econe}). Indeed, the linearity property of weak measurements yields
a ``geometrical picture'' for weak values of  spin
components of a spin-$1\over 2$ particle. The operators $\sigma_x$,
$\sigma_y$, and $\sigma_z$ are a complete set of spin operators and they
yield a geometry in the familiar three-dimensional space. Each
generalized two-state vector of a spin-$1\over 2$ particle corresponds
to a vector in this three-dimensional space with components equal to
the weak values of $\sigma_x$,
$\sigma_y$, and $\sigma_z$. We call it ``weak vector''. The weak value
of a spin component in an arbitrary direction, then, is given by the
projection of the weak vector on this direction. If the weak vector is
real and its value larger than 1, then there is a cone of directions
the projection on which is equal 1. This yields an alternative proof
that in some situations there is a continuum of directions forming a
cone in which the result of a spin-component measurements are known
with certainty, see \ref{sec:imgs}.

\subsection{
 Experimental realizations of weak measurements}

Realistic weak measurements (on an ensemble) involve preparation of
a large pre-selection ensemble, coupling to the measuring devices of
each element of the ensemble, post-selection measurement which, in
all interesting cases, selects only a small fraction of the original
ensemble, selection of corresponding measuring devices, and
statistical analysis of their outcomes. In order to obtain good
precision, this selected ensemble of the measuring devices has to be
sufficiently large.  Although there are significant technological
developments in ``marking'' particles running in an experiment,
clearly the most effective solution is that the particles themselves
serve as measuring devices. The information about the measured
variable is stored, after the weak measuring interaction, in their
other degree of freedom. In this case, the post-selection of the
required final state of the particles automatically yields the
selection of the corresponding measuring devices . The requirement
for the post-selection measurement is, then, that there is no
coupling between the variable in which the result of the weak
measurement is stored and the post-selection device.

 An example of such a case is the Stern-Gerlach experiment where the
 shift in the momentum of a particle, translated into a spatial shift,
 yields the outcome of the spin measurement. Post-selection
 measurement of a spin component in a certain direction can be
 implemented by another (this time strong) Stern-Gerlach coupling
 which splits the beam of the particles. The beam corresponding to the
 desired value of the spin is then analyzed for the result of the weak
 measurement. The requirement of non-disturbance of the results of the
 weak measurement by post-selection can be fulfilled by arranging the
 shifts due to the two Stern-Gerlach devices to be orthogonal to each
 other. The details are spelled out in \cite{s-100}.

 An analysis of a realistic experiment which can yield large weak value
 $Q_w$ appears in \cite{KV}. Duck, Stevenson, and Sudarshan \cite{duck}
 proposed a slightly different optical realization which uses
 a birefringent plate instead of a prism. In this case the measured
 information is stored directly in the spatial shift of the beam
 without being generated by the shift in the momentum.  Ritchie,
 Story, and Hulet adopted this scheme and performed the first
 successful experiment measuring the weak value of the polarization
 operator \cite{RIT}. Their results are in very good agreement with
 theoretical predictions.  They obtained weak values which are very
 far from the range of the eigenvalues, ($-1, 1$), their highest
 reported result is $Q_w = 100$. The discrepancy between calculated
 and observed weak value was 1\%. The RMS deviation from the mean of
 16 trials was 4.7\%. The width of the probability distribution was
 $\Delta = 1000$ and the number of pre- and post-selected photons was
 $N \sim 10^{8}$, so the theoretical and experimental uncertainties
 were of the same order of magnitude. Their other run, for which they
 showed experimental data on graphs (which fitted very nicely
 theoretical graphs), has the following characteristics: $Q_w = 31.6$,
 discrepancy with calculated value 4\%, the RMS deviation 16\%,
 $\Delta = 100$, $N \sim 10^5$.
Similar  optical experiment  has been successfully performed several
years ago \cite{PARK}.

 Recently, optical weak measurement
experiments moved to the field of fiber optics
\cite{fiop1,fiop2,fiop3}. Another step prevents now any sceptic to
argue that the unusual outcomes of weak measurement is a classical
effect because macroscopic number of photons are involved in these
experiments. The weak measurement of photon polarization have been
performed with single particles \cite{1phwmpol}. Note also a more
controversial issue of measurement of ``time of arrival''
\cite{arr0} for which weak measurement technique were also applied
\cite{arr1,arr2,arr3}.

Already at 1990 \cite{AV90} we gave an example of a gedanken
experiment in which pre- and post-selection lead to a superluminal
propagation of light. Steinberg and Chiao \cite{Stein-lev,Chiao}
connected this to superluminal effect observed for tunneling
particles. The issue was analyzed recently by Aharonov et al.
\cite{tunA} and Sokolovsky et al. \cite{tunS}. Rohrlich and Aharonov
also predicted that there is really a physical meaning for this
superluminal propagation: we should expect Cherenkov radiation in
such experiment \cite{cher}.

Note also proposals for {\it weak nonlocal} measurements
\cite{weanl1,weanl2}. In this works it was pointed out that
observation of correlations in between outcomes of local weak
measurements can yield values of nonlocal variables. However, these
methods are very inefficient, and the methods of efficient nonlocal
measurements \cite{AAV86} require conditions which contradict
conditions of weak measurements, so we doubt that there will be
efficient weak nonlocal measurement proposals suitable for
realization in a laboratory.

\section{The quantum time-translation machine}

\subsection{Introduction}

To avoid possible misinterpretations due to the name ``time machine,''
let us explain from the outset what our machine \cite{t-m} can do and how it
differs from the familiar concept of ``time machine.'' Our device is not
for time travel. All that it can accomplish is to change the rate of
time flow for a closed quantum system.
Classically, one can slow down the time flow of
a system relative to an external observer, for example, by fast
travel. Our quantum time machine is able to change the rate of time
flow of a system for a given period by an arbitrary, even {\it negative},
factor. Therefore, our machine, contrary to any classical device, is
capable of moving the system to its ``past.'' In that case, at the
moment the machine completes its operation the system is in a state
in which it was some time {\it before the beginning} of the operation of the
time machine. Our machine can also move the system to the future,
i.e., at the end of the operation of the time machine the system is in
a state corresponding to some later time of the undisturbed evolution.

A central role in the operation of our time machine is played by a peculiar
mathematical identity which we discuss in
Sec. \ref{sec:iden}.
In order to obtain different time evolutions of the system we use the gravitational time dilation effect which is
discussed in Sec. \ref{sec:ctm}.
 In
Sec. \ref{sec:qgtm} we describe the design and the operation of our time machine.
 The success of the operation of our time
machine depends on obtaining a specific outcome in the post-selection
quantum measurement. The probability of the successful post-selection
measurement is analyzed in Sec. \ref{sec:probsuc}.
The  concluding discussion of the limitations and the advantages of our time
machine appear in Sec. \ref{sec:pastfutu}.

\subsection{A peculiar mathematical identity}
\label{sec:iden}

The peculiar interference effect of weak measurements (\ref{sup1}),
that a particular
superposition of identical Gaussians shifted by small values yields
the Gaussian shifted by a large value occurs not just for Gaussians,
but for a large class of functions. Consider now that the system is
described by such a wave function and the shifts are due to the time
evolutions for various periods of time. Then, this effect can be a
basis of a (gedanken) time-machine. A specific
superposition of time evolutions for short periods of time $\delta t_n$
yields a time evolution for a large period of time $\Delta t$
\begin{equation}
\label{suptime}
\sum_{n=0}^N \alpha_n U(\delta t_n) |\Psi \rangle
\sim  U(\Delta t) |\Psi \rangle .
\end{equation}
This approximate equality holds (with the same $\delta t_n$ and
$\Delta t$) for a
large class of states $ |\Psi \rangle$ of the quantum system, and in some cases
even for all states of the system.

In order to obtain different time evolutions $U(\delta t_n)$ we use the
gravitational time dilation effect. For finding the appropriate
$\delta t_n$ and $\alpha_n$ we will rely on the identity (\ref{sup1}) for a
particular weak measurement. We choose
\begin{equation}
  \label{cnan}
  c_n =n/N, ~~~~~\alpha_n = {{N!}\over {(N-n)! n!}}\eta^n (1-\eta)^{N-n} ,
\end{equation}
where $ n =0,1,..., N$. Note, that the coefficients $\alpha_n$ are
terms in the binomial expansion of $[\eta + (1- \eta)]^N$ and, in
particular, $\sum_{n=0}^N \alpha_n =1$. The corresponding ``weak
  value'' in this case is $\eta$ and for a large class of functions
  (the functions with Fourier transform bounded by an exponential) we
  have an approximate equality
\begin{equation}
\label{sup2}
 \sum_{n=0}^N  \alpha_n f(t-c_n) \simeq  f(t-\eta) .
\end{equation}
The proof can be found in Ref. \cite{tm-va}. Here we only demonstrate
it on a numerical example, Fig. 5. Even for a relatively small number
of terms in the sum (14 in our example), the method works remarkably
  well. The shifts from 0 to 1 yield the shift by 10. The distortion
  of the shifted function is not very large. By increasing the number
  of terms in the sum the distortion of the shifted function can be
  made arbitrarily small.

\vskip .5cm
\epsfxsize=10cm
\centerline{\epsfbox{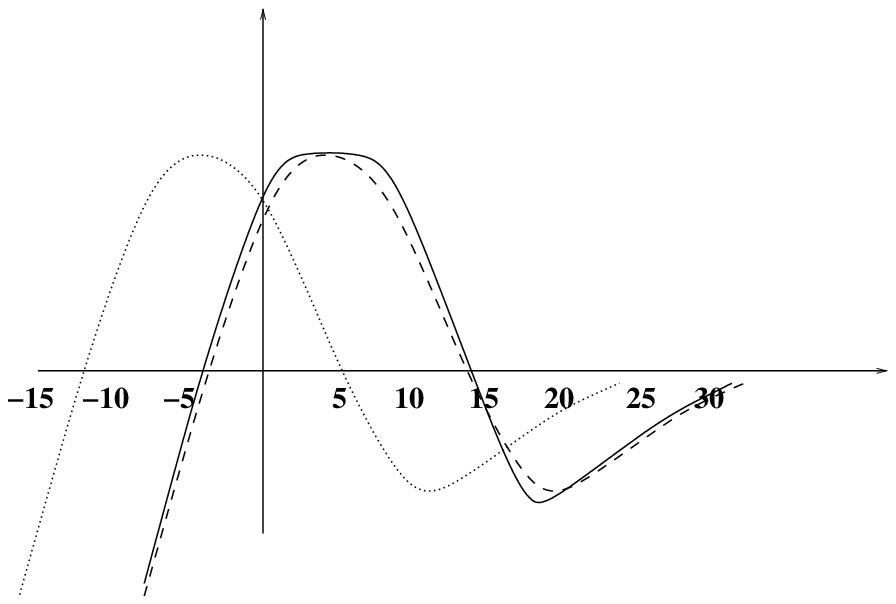}}
\vskip .2cm
\noindent

{\small{\bf Fig. 5. Demonstration of an approximate equality given by
(\ref{sup2}).} The sum of a function shifted by the 14 values $c_n$
between 0 and 1 and multiplied by the coefficients
 $\alpha_n$ ($c_n$ and $\alpha_n$ are
given by (\ref{cnan}) with $N=13$, $\eta=10$) yields approximately the
same function shifted by the value 10.} The dotted line shows $f(t)$;
the dashed line shows$f(t-10)$, the RHS of (\ref{sup2});
and the solid line shows the sum, the  LHS of (\ref{sup2}).

\subsection{Classical time machines}
\label{sec:ctm}

A well-known example of a time machine is a rocket which takes a
system to a fast journey. If the rocket is moving with velocity $V$ and
the duration of the journey (in the laboratory frame) is $T$, then we
obtain the time shift (relative to the situation without the fast
journey):

\begin{equation}
  \label{dtv}
  \delta t =T\left(1- \sqrt{1-{V^2\over c^2}}\right) .
\end{equation}
For typical laboratory velocities
this effect is rather small, but it has been observed experimentally
in precision measurements in satellites and, of course, the effect is observed on decaying particles in accelerators. In such a ``time machine,"
however, the system necessarily experiences external force, and we
consider this a conceptual disadvantage.

  In our time machine we use,
instead of the time dilation of special relativity, the gravitational
time dilation. The relation between the proper time of the system
placed in a gravitational potential $\phi$  and the time of the external
observer ( $\phi  = 0$) is given by $d\tau = dt \sqrt{1 + 2\phi/c^2}$.
We produce the
gravitational potential by surrounding our system with a spherical
shell of mass $M$ and radius $R$. The gravitational potential inside the
shell is $\phi = -GM/R$. Therefore, the time shift due to the massive
shell surrounding our system, i.e., the difference between the time
period $T$ of the external observer at a large distance from the shell
and the period of the time evolution of the system (the proper time),
is
\begin{equation}
  \label{dtg}
  \delta t =T\left(1- \sqrt{1-{2GM\over {c^2 R}}}\right) .
\end{equation}
 This effect, for any man-made massive
shell, is too small to be observed by today's instruments. However,
the conceptual advantage of this method is that we do not ``touch'' our
system. Even the gravitational field due to the  massive spherical
shell vanishes inside the shell.

 The classical time machine can only
{\it slow down} the time evolution of the  system. For any reasonable mass
and radius of the shell the change of the rate of the time flow is
extremely small.  In the next section we shall describe our quantum
time machine which amplifies the effect of the classical
gravitational time machine (for a spherical shell of the same mass),
and  makes it possible to speed up the time flow for an evolution of
a system, as well as to change its direction.

\subsection{ Quantum gravitational time machine }
\label{sec:qgtm}

In our machine we use the gravitational time dilation and a quantum
interference phenomenon which, due to the peculiar mathematical
property discussed in Sec.\ref{sec:iden}, amplifies the time translation. We
produce the superposition of states shifted in time by small
values   $\delta t_n$ (due to spherical shells of different radii) given by
the left-hand side of (\ref{suptime}). Thus,
we obtain a time shift by a possibly large, positive or negative, time
interval $\Delta t$.

The wave function of a quantum system $\Psi(q, t)$, considered as a
function of time, usually has a Fourier transform which decreases
rapidly for large frequencies. Therefore, the sum of the wave functions
shifted by small periods of time $\delta t_n =\delta t c_n$, and
multiplied by the coefficients $\alpha_n$, with $c_n$ and $\alpha_n$
given by (\ref{cnan}), is approximately equal to the wave function
shifted by the large time $\Delta t = \delta t \eta$.
Since the equality (\ref{sup2})
is correct with the same coefficients for all functions with rapidly
decreasing Fourier transforms, we obtain for each $q$, and therefore for
the whole wave function,
\begin{equation}
\label{sup3}
 \sum_{n=0}^N  \alpha_n \Psi (q,t - \delta t_n)
  \simeq  \Psi (q, t - \Delta t) .
\end{equation}
 Thus, a device which changes the state of the
system from $\Psi(q, t)$ to the state given by the
 left-hand side of (\ref{sup3}) generates the time shift of $\Delta t$.
 Let us now present a
design for such a device and explain how it operates.

 Our machine
consists of the following parts: a massive spherical shell, a
mechanical device - ``the mover'' - with a quantum operating system, and a
measuring device which can prepare and verify states of this quantum
operating system.

 {\it The massive shell} of mass $M$ surrounds our system
and its radius $R$ can have any of the values $R_0, R_1 ,..., R_N$.
Initially, $R = R_0$.

{\it  The mover} changes the radius of the spherical
shell at time $t = 0$, waits for an (external) time T, and then moves it
back to its original state, i.e., to the radius $R_0$.

{\it The quantum
operating system} (QOS) of the mover controls the radius to which the
shell is moved for the period of time $T$. The Hamiltonian of the QOS
has $N + 1$ non-degenerate eigenstates $|n \rangle$, $n = 0, 1,..., N$. If the
state of the QOS is $|n \rangle$, then the mover changes the radius of the
shell to the value $R_n$.

{\it The measuring device} preselects and
post-selects the state of the QOS. It prepares the QOS before the time
$t = 0$ in the initial state
\begin{equation}
  \label{psiqos}
  |\Psi_{in}\rangle_{QOS} = {\cal N}  \sum_{n=0}^N  \alpha_n  |n \rangle ,
\end{equation}
with the normalization factor
\begin{equation}
  \label{norm}
   {\cal N} = {1\over \sqrt{  \sum_{n=0}^N | \alpha_n|^2}} .
\end{equation}

After the mover completes its operation, i.e., after the time $t = T$,
we perform another measurement on the QOS. One of the
non-degenerate eigenstates of this measurement is the specific ``final
state''
\begin{equation}
  \label{psiqosf}
  |\Psi_{f}\rangle_{QOS} ={1\over \sqrt{N+1}}  \sum_{n=0}^N    |n \rangle.
\end{equation}
Our machine works only if the post-selection measurement yields the
state (\ref{psiqosf}). Unfortunately, this is a very rare event. We shall
discuss the probability of obtaining the appropriate outcome in the
next section.

Assume that the post-selection measurement is successful, i.e., that we
do obtain the final state (\ref{psiqosf}). We will next show that in
this case, assuming an appropriate choice of the radii $R_n$, our
``time machine'' shifts the wave function of the system by the time
interval $\Delta t$.  The time shift is defined relative to the
situation in which the machine has not operated, i.e., the radius of
the shell was not changed from the initial value $R_0$. In order to
obtain the desired time shift $\Delta t = \delta t \eta$ we chose the
radii $R_n$ such that
\begin{equation}
  \label{dtn}
\delta t_n
\equiv {{n  \delta t}\over N} = T\left (\sqrt{1-{2GM\over {c^2 R_0}}}
-\sqrt{1-{2GM\over {c^2 R_n}}}\right ) .
\end{equation}
The maximal time shift in the different terms of the superposition
 (the left-hand side of (\ref{suptime})) is $\delta t_N = \delta t$. The
parameter $\eta$ is the measure of the ``quantum amplification''
relative to the maximal (classical) time shift $\delta t$. If the
radius $R_0$ of the shell is large enough that the time dilation due
to the shell in its initial configuration can be neglected, (\ref{dtn})
simplifies to
\begin{equation}
  \label{dtns}
 \delta t_n  = T \left (1-\sqrt{1-{2GM\over {c^2 R_n}}}\right ) .
\end{equation}

Let us assume then that we have arranged the radii according to Eq.
(\ref{dtns}) and we have prepared the quantum operating system of the
mover in the state (\ref{psiqos}). Then, just prior to the operation of
the time
machine the overall state is the direct product of the corresponding
states of the system, the shell, and the mover,
\begin{equation}
\label{psiall1}
{\cal N}  |\Psi (q,0)\rangle |R_0\rangle \sum_{n=0}^N  \alpha_n  |n \rangle ,
\end{equation}
where  $|R_0\rangle$ signifies that the shell, together with the mechanical
part
of the mover, is at the radius $R_0$. Although these are clearly
macroscopic bodies. we assume that we can treat them
quantum-mechanically. We also make an idealized assumption that these
bodies do not interact with the environment, i.e.. no element of the
environment becomes correlated to the radius of the shell.

Once the mover has operated, changing the radius of the spherical
shell, the overall state becomes
\begin{equation}
  \label{psiall2}
{\cal N}  |\Psi (q,0)\rangle  \sum_{n=0}^N  \alpha_n |R_n\rangle |n \rangle .
\end{equation}

For different radii $R_n$, we have different gravitational potentials
inside the shell and, therefore, different relations between the flow
of the proper time of the system and the flow of the external time.
Thus, after the external time $T$ has elapsed, just before the mover
takes the radii $R_n$ back to the value $R_0$, the overall state is
\begin{equation}
  \label{psiall3}
  {\cal N}   \sum_{n=0}^N  \alpha_n|\Psi (q,T - \delta t_n)\rangle
 |R_n\rangle |n \rangle .
\end{equation}
Note that now the system, the
shell, and the QOS are correlated: the system is not in a pure quantum
state. After the mover completes its operation, the overall state
becomes
\begin{equation}
  \label{psiall4}
  {\cal N}   \sum_{n=0}^N  \alpha_n|\Psi (q,T - \delta t_n)\rangle
 |R_0\rangle |n \rangle ,
\end{equation}
 There is still a
correlation between the system and the QOS.

The last stage is the post-selection
measurement performed on the QOS. It puts the QOS and, consequently,
our quantum system, in a pure state. After the successful
post-selection measurement, the overall state is
\begin{equation}
  \label{psiall5}
  \left ( \sum_{n=0}^N  \alpha_n|\Psi (q,T - \delta t_n)\rangle \right )
 |R_0\rangle \left ( {1\over \sqrt{N+1}}  \sum_{n=0}^N    |n \rangle \right ) .
\end{equation}

We have shown that the wave function of the quantum system $\Psi
(q,t)$ is changed by the operation of the time machine into $
\sum_{n=0}^N \alpha_n|\Psi (q,T - \delta t_n)$. Up to the precision of
the approximate equality (\ref{sup2}) (which can be arbitrarily
improved by increasing the number of terms $N$ in the sum), this wave
function is indeed $|\Psi (q,T - \Delta t)\rangle$! Note that for
$\Delta t > T$, the state of the system at the
 moment the time machine has
completed its operation is the state in which the system was before
the beginning of the operation of the time machine.

\subsection{The probability of the success of the quantum time machine}
\label{sec:probsuc}

The main conceptual weakness of our time machine is that usually it
does not work. Successful post-selection measurements corresponding to
large time shifts are extremely rare. Let us estimate the probability
of the successful post-selection measurement in our example. The
probability is given by the square of the norm of the vector obtained
by projecting the state (\ref{psiall5}) on the subspace defined by state
(\ref{psiqosf}) of the QOS:
\begin{equation}
  \label{Prob1}
  {\rm Prob} = || {{\cal N}\over \sqrt{N+1}}
  ( \sum_{n=0}^N  \alpha_n|\Psi (q,T - \delta t_n)\rangle
 |R_0\rangle  ||^2 .
\end{equation}

In order to obtain a time shift without significant distortion,
the wave functions shifted by different times  $\delta t_n$  have to be such
that the scalar products
between them can be approximated by 1.
Taking then the explicit
form of $\alpha_n$  from (\ref{cnan}), we evaluate the
probability (\ref{Prob1}), obtaining
\begin{equation}
  \label{Prob23}
  {\rm Prob} \simeq {{\cal N}^2\over N} .
\end{equation}
The normalization factor $\cal N$ given by (\ref{norm}) decreases very
rapidly for large $N$. Even if we use a more efficient choice of the
initial and the final states of the QOS (see Ref. \cite{AV90})
for the amplification, $\eta >1$,  the probability
decreases with $N$ as $1/(2\eta -1 )^N$.

 The small probability of the successful operation of our time machine
  is, in fact, unavoidable. At the time just before the
 post-selection measurement, the system is in a mixture of states
 correlated to the orthogonal states of the QOS [see (\ref{psiall4})].
 The probability of finding the system at that time in the state
 ${|\Psi (q,T - \Delta t)\rangle}$, for $\Delta t$ which differs significantly
 from the time periods $\delta t_n$, is usually extremely small. This
 is the probability to find the system, by a measurement performed
 ``now,'' in the state in which it was supposed to be at some other
 time. For any real situation this probability is tiny but not
 equal precisely to zero, since all systems with bounded energies have
 wave functions with non-vanishing tails. The  successful operation
 of our time machine is a particular way of ``finding''  the state of
 the quantum system shifted by the period of time $\Delta  = \eta \delta$.
 Therefore, the probability for success cannot be larger than the
 probability  of finding the shifted wave function by direct
 measurement.

 One can wonder what has been achieved by all this rather complicated
 procedure if we can obtain the wave function of the system shifted by
 the time period $\Delta t$ simply by performing a quantum
 verification measurement at the time $T$ of the state $|\Psi (q,T -
 \Delta t)\rangle$. There is a very small chance for the success of
 this verification measurement, but using our procedure the chance is
 even smaller. What our machine can do, and we are not aware of any
 other method which can achieve this, is to shift the wave function in
 time {\it without knowing} the wave function. If we obtain the
 desired result of the post-selection measurement (the post-selection
 measurement performed on {\it the measuring device}), we know that
 the wave function of the system, whatever it is, is shifted by the
 time $\Delta t$. Not only is the knowledge of the wave function of
 the system inessential for our method, but even the very nature of
 the physical system whose wave function is shifted by our time
 machine need not be known. The only requirement is that the energy
 distribution of the system decreases rapidly enough.  If the
 expectation value of the energy can be estimated, then we can improve
 dramatically the probability of the success of our procedure. The
 level of difficulty of the time shift without distortion depends on
 the magnitude of the energy dispersion $\Delta E$ and not on the
 expectation value of energy $\langle E\rangle$.  For quantitative
 analysis of this requirement see \cite{tm-va}.

 The operation of our time machine can be considered as {\it a
   superposition of time evolutions} \cite{t-m} for different periods
 of time $\delta t_n$. This name is especially appropriate if the
 Hamiltonian of the system is bounded, since in this case the
approximate equality (\ref{suptime})
is correct for all states $|\Psi \rangle$.

\subsection{Time translation to the past and to
                   the future}
\label{sec:pastfutu}

Let us spell out again what our machine does. Assume that the time
evolution of the state of the system is given by $|\Psi (t)\rangle$.
By this we mean that this is the evolution {\it before} the operation
of the time machine and this is also the evolution later, provided we
{\it do not} operate the time machine. The state $|\Psi (t)\rangle$
describes the actual past states of the system and the counterfactual
future states of the system, i.e., the states which will be in the
case we do not disturb the evolution of the system by the operation of
our time machine. Define ``now,'' $t = 0$, to be the time at which we
begin the operation of the time machine. The time interval of the
operation of the time machine is $T$. Moving the system to the {\it
  past} means moving it to the state in which the system actually was
at some time $t < 0$. Moving the system to the future means moving it
to the state in which it would have wound up after undisturbed
evolution at some future time $t > T$.  Evidently, the classical time
machine does neither of these, since all it can achieve is that at
time $T$ the system is in the state corresponding to the time
$t$, $0 < t < T$.

 When we speed up or slow down the rate of the time
evolution, the system passes through all states of its undisturbed
evolution only once. More bizarre is the situation when we reverse the
direction of the time flow, thus ending up, after completing the
operation of the time machine, in the state in which the system was
before $t = 0$. In this case the system passes three times through some
states during its evolution.

 For our time machine to operate
properly, it is essential that the system is isolated from the
external world. In the case of the time translation to the
state of the past, the system has to be isolated not only during the time of
the operation of the time machine, but also during the whole period of
intended time translation. If the system is to be moved to the state
in which  it was at the time $t$, $t < 0$, then it has to be isolated
from the time $t$ until  the end  of operation of the time machine.
This seems to be a limitation
 of our time machine. It leads,
however, to an interesting possibility. We can send a system to its
{\it counterfactual past}, i.e., to the past in which it was supposed
to be if it were isolated (or if it were in any environment chosen by
us).

 Consider an excited atom which we isolate in the vacuum at time $t = 0$ inside our time machine. And assume that our time machine made a
successful time translation to a negative time $t$, such that $|t|$ is
larger than the lifetime of the excited atomic state. Since the atom,
now, is not in the environment it was in the past, we do not move the
atom to its actual state in the past. Instead, we move the atom to the
state of its counterfactual past.  By this we mean the state of the
isolated atom which, under its normal evolution in the vacuum during
the time period $|t|$ winds up in the excited state. In fact, this is
the state of the atom together with an incoming radiation field. The
radiation field is exactly such that it will be absorbed by the atom.
Although our procedure is very complicated and only very rarely
successful, still, it is probably the easiest way to prepare the
precise incoming electromagnetic wave which excites a single atom
with probability one.

\subsection{Experimental realization of the quantum time-translation machine?!}
\label{sec:exptm}

Suter \cite{SUT} has claimed to perform an experimental realization
of the quantum time-translation machine using a classical Mach-Zehnder
interferometer.  The experimental setup of Suter, however, does not
fall even close to the definition of the time machine. In his setup we
know what is the system and what is its initial state. What he shows
is that if we send a single mode of a radiation field through a
birefringent retardation device which yields different retardations
for two orthogonal polarizations, then placing the pre-selection
polarization filter and the post-selection polarization filter will
lead to a much larger effect than can be achieved by  pre-selection alone. Thus, it might seem like speeding up the time evolution,
but this procedure fails all tests of universality.  Different modes
of radiation field speed up differently, an arbitrary wave packet is
usually distorted, and for other systems (other particles) the device
is not supposed to work at all.

Thus, the first basic requirement that the time machine has to work
for various systems is not fulfilled from the beginning. And it cannot
be easily modified since the ``external'' variable (which is supposed
to be a part of the time-machine) is the property of the system itself
-- the polarization of the radiation field.  The next necessary
requirement, that it works for a large class of the initial states of
the system, cannot be fulfilled too.  Indeed, he considers a
superposition of only two time evolutions.  This superposition can be
identical to a longer evolution for a particular state, but not for a
large class of states. As it has been shown \cite{t-m,tm-va} a
superposition of a large number of time evolutions is necessary for
this purpose.

Suter, together with R. Ernst and M. Ernst, performed in
the past another experiment which they called ``An experimental
realization of a quantum time-translation machine'' \cite{SEE}. In
this experiment a very different system was used: the effect was
demonstrated on the heteronuclear coupling between two nuclear spins.
But the experimental setup was also applicable only to a specific
system and only for a certain state. Therefore, the same criticism is
applicable and, therefore, one should not call it an implementation of
the time-translation machine.

Although the experiments of Suter are not implementations of the
quantum time machine, still, they are interesting as {\em weak
  measurements}.  The experiment of Suter with a birefringent
retardation device can be considered as a weak measurement of a
polarization operator. In fact, this is a variation of the experiments
which were proposed \cite{KV} and performed \cite{RIT} previously.
The ``weakness condition'' of these two experiments follows from the
localization of the beam (which was sent through a narrow slit).  The
``weak'' regime of the experiment of Suter is achieved by taking the
retardation small.  The second experiment of Suter can be considered
as the first weak measurement of a nuclear spin component.

\section{Time Symmetry}

\subsection{Forward and Backward Evolving Quantum States}
\label{sec:fbeqs}

Before discussing  the time symmetry of the pre- and post-selected
systems which are usually discussed in the framework of the
two-state vector formalism, we will consider the question of
differences between possibilities for  manipulating  forward
evolving quantum states (\ref{1qs}) and backward evolving state
(\ref{bra}) which has been recently analyzed \cite{backV}. It is
particularly important in the light of recent argument of Shimony
\cite{shi-last} against equal status of forward and backward
evolving quantum states.

A notable difference between forward and backward evolving states
has to do with the {\it creation} of a particular quantum state at a
particular time. In order to create the quantum state $|A=a\rangle$
evolving forward in time, we measure $A$ before this time. We cannot
be sure to obtain $A=a$, but if we obtain a different result $A=a'$
we can always perform a unitary operation and thus create at time
$t$ the state $|A=a\rangle$. On the other hand, in order to create
the backward evolving quantum state $\langle A=a|$, we measure $A$
after time $t$. If we do not obtain the outcome $A=a$, we cannot
repair the situation, since the correcting transformation has to be
performed at a time when we do not yet know which correction is
required. Therefore, a backward evolving quantum state at a
particular time can be created only with some probability, while a
forward evolving quantum state can be created with certainty. (Only
if the forward evolving quantum state is identical to the backward
evolving state we want to create at time $t$, and only if we know
that no one touches the system at time $t$, can the backward
evolving state be created with certainty, since then the outcome
$A=a$ occurs with certainty. But this is not an interesting case.)

The formalism of quantum theory is time reversal invariant. It does
not have an intrinsic arrow of time. The difference with regard to
the creation of backward and forward evolving quantum state follows
from the ``memory's'' arrow of time. We can base our decision of
what to do at a particular time only on events in the past, since
future events are unknown to us. The memory time arrow is
responsible for the difference in our ability to manipulate forward
and backward evolving quantum states. However, the difference is
only in relation to creation of the quantum state. As we will see
below there are no differences with measurements in the sense of
``finding out'' what is the state at a particular time.

 The  ideal (von Neumann) measurement procedure applies both to
forward evolving quantum states and to backward evolving quantum
states. In both cases, the outcome of the measurement is known after
the time of the measurement. All that is known about what can be
measured in an ideal (nondemolition) measurement of a forward
evolving quantum state can be applied also to a backward evolving
quantum state. There are constraints on the measurability of
nonlocal variables, i.e. variables of composite systems with parts
separated in space. When we consider instantaneous nondemolition
measurements (i.e. measurements in which, in a particular Lorentz
frame during an arbitrarily short time, local records appear which,
when taken together, specify the eigenvalue of the nonlocal
variable), we have classes of measurable and unmeasurable variables.
For example, the Bell operator variable is measurable, while some
other variables \cite{VP}, including certain variables with product
state eigenstates \cite{NLWE,NLWEVG}, cannot be measured.

The procedure for measuring nonlocal variables involves entangled
ancillary particles and local measurements, and can get quite
complicated. Fortunately, there is no need to go into detail in
order to show the similarity of the results for forward and backward
evolving quantum states. The operational meaning of the statement
that a particular variable $A$ is measurable is that in a sequence
of three consecutive measurements of $A$ - the first taking a long
time and possibly including bringing separate parts of the system to
the same location and then returning them, the second being short
and nonlocal, and the third, like the first, consisting of bringing
together the parts of the system - all outcomes have to be the same.
But this is a time symmetric statement; if it is true, it means that
the variable $A$ is measurable both for forward and backward
evolving quantum states.

We need also to obtain the correct probabilities in the case that
different variables are measured at different times. For a forward
evolving quantum state it follows directly from the linearity of
quantum mechanics. For a backward evolving quantum state, the
simplest argument is the consistency between the probability of the
final measurement, which is now $B=b$, given the result of the
intermediate measurement $A=a$, and the result of the intermediate
measurement given the result of the final measurement. We assume
that the past is erased. The expression for the former is $|~
\langle A=a |B=b\rangle|^2$. For consistency, the expression for the
latter must be the same, but this is what we need to prove.

In exactly the same way we can show that the same procedure for {\it
teleportation} of a forward evolving quantum state \cite{tele}
yields also teleportation of a backward evolving quantum state. As
the forward evolving quantum state is teleported to a space-time
point in the future light cone, the backward evolving quantum state
is teleported to a point in the backward light cone. Indeed, the
operational meaning of teleportation is that the outcome of a
measurement in one place is invariably equal to the outcome of the
same measurement in the other place. Thus, the procedure for
teleportation of the forward evolving state to a point in the future
light cone invariably yields teleportation of the backward evolving
quantum state to the backward light cone.

The impossibility of teleportation of the backward evolving quantum
state outside the backward light cone follows from the fact that it
will lead to teleportation of the forward evolving quantum state
outside the forward light cone, and this is impossible since it
obviously breaks causality.

Another result which has been proved using causality argument is the
{\it no cloning theorem} for backward evolving quantum quantum
states \cite{backV}. So, also in this respect there is no difference
between forward and backward evolving quantum states.

The argument used above does not answer the question of whether it
is possible to measure nonlocal variables in a {\it demolition
measurement}. Demolition measurements destroy (for the future) the
state and may be the quantum systems itself. Thus, obviously, a
demolition measurement of a nonlocal variable of a quantum state
evolving forward in time does not measure this variable for a
quantum state evolving backward in time. Any nonlocal variable of a
composite system can be measured with demolition for a quantum state
evolving forward in time \cite{V}. Recently, it has been shown
\cite{VN} also that any nonlocal variable can be measured for a
quantum state evolving backward in time. Moreover, the procedure is
simpler and requires fewer entanglement resources.

The difference follows from the fact that we can change the
direction of time evolution of a backward evolving state along with
complex conjugation of the quantum wave (flipping a spin). Indeed,
all we need is to prepare an EPR state of our system and an ancilla.
Guarding the system and the ancilla ensures that the forward
evolving quantum state of the ancilla is the flipped state of the
system. For a spin wave function we obtain
 $\alpha \langle{\uparrow}
|+\beta\langle{\downarrow}| \rightarrow
   -\beta^\ast |{\uparrow}\rangle +\alpha ^*|{\downarrow}\rangle $.
For a continuous variable wave function $\Psi (q)$ we need the
original EPR state $|q-\tilde{q}=0, ~~p+\tilde{p}=0\rangle$. Then,
the backward evolving quantum state of the particle will transform
into a complex conjugate state of the ancilla $ \Psi(q) \rightarrow
\Psi^\ast(\tilde{q})$.

If the particle and the ancilla are located in different locations,
then such an operation is a combination of time reversal and
teleportation of a backward evolving quantum state of a continuous
variable \cite{V94}.

We cannot flip and change the direction of time evolution of a
quantum state evolving forward in time. To this end we would have to
perform a Bell measurement on the system and the ancilla and to get
a particular result (singlet). However, we cannot ensure this
outcome, nor can we correct the situation otherwise. Moreover, it is
easily proven that no other method will work either. If one could
have a machine which turns the time direction (and flips) a forward
evolving quantum state, then one could prepare at will any state
that evolves toward the past, thus signalling to the past and
contradicting causality.

Let us consider now a pre- and post-selected system. It is
meaningless to ask whether we can perform a nondemolition
measurement on a system described by a two-state vector. Indeed, the
vector describing the system should not be changed {\it after} the
measurement, but there is no such time: for a forward evolving
state, ``after'' means later, whereas for a backward evolving state,
``after'' means before. It is meaningful to ask whether we can
perform a {\it demolition} measurement on a system described by a
two-state vector. The answer is positive \cite{VN}, even for
composite systems with separated parts.

Next, is it possible to teleport a two-state vector? Although we can
teleport both forward evolving and backward evolving quantum states,
we cannot teleport the two-state vector. The reason is that the
forward evolving state can be teleported only to the future light
cone, while the backward evolving state can be teleported only to
the backward light cone. Thus, there is no space-time point to which
both states can be teleported.

Finally, the answer to the question of whether it is possible to
clone a two-state vector is negative, since neither forward evolving
nor backward evolving quantum states can be cloned.

\subsection{Time-symmetric aspects of pre- and post-selected systems}
\label{sec:tsa}

 When a quantum system is described by the
two-state vector (\ref{2sv}) or the generalized two-state vector
(\ref{g2sv}), the backward-evolving states enter on equal footing
with   the forward-evolving states.  Note that the asymmetry in  the
procedure for obtaining the state (\ref{g2sv}) is not essential: we
can start preparing  $1/\sqrt N \sum_i~ |\Phi_i \rangle |i \rangle $
instead.

We will  analyze now the symmetry under the interchange
$\langle\Phi|~|\Psi \rangle ~\leftrightarrow ~ \langle \Psi|~|\Phi
\rangle$.  This will be considered as a symmetry under reversal of the
direction of the arrow of time.  It is important to note that in
general this interchange is not equivalent to the interchange of  the
measurements creating the two-state vector $A=a$ and  $B=b$.
 An example showing the non-equivalence can be found in
 \cite{SHI}.  However, in order to simplify the
discussion, we will assume that the free Hamiltonian is zero, and
therefore $|\Psi \rangle ~=~ |A=a\rangle$ and $\langle\Phi |
~=~\langle B=b|$. In this case, of course, the reversal of time arrow
is identical to the interchange of the measurements at $t_1$ and
$t_2$. If the free Hamiltonian is  not zero, then an
appropriate modification  should be made  \cite{va-tsqt}.

The ABL rule for the probabilities of the outcomes of ideal
measurements (\ref{ABL}) is also explicitly time-symmetric: First,
both $\langle\Phi |$ and $|\Psi \rangle$ enter the equation on equal
footing. Second, the probability (\ref{ABL}) is unchanged under the
interchange $\langle\Phi|~|\Psi \rangle~ \leftrightarrow ~\langle
\Psi|~|\Phi \rangle$.

 The ABL rule for  a quantum system described by a  generalized two-state
vector (\ref{g2sv}) is time-symmetric as well:
  $\langle\Phi_i |$ and ~$|\Psi_i \rangle$ enter the equation on equal
 footing. The manifestation of the symmetry of this formula under the
 reversal of the arrow of time includes complex conjugation of the
 coefficients. The
 probability (\ref{ABL-gen})  is unchanged  under the interchange
$\sum_i \alpha_i \langle \Phi_i | ~ | \Psi_i \rangle ~ \leftrightarrow
~ \sum_i \alpha_i^{\ast} \langle \Psi_i | ~ | \Phi_i \rangle$.

The outcomes of weak measurements, the weak values, are also symmetric under the interchange  $\langle\Phi |~
|\Psi \rangle ~ \leftrightarrow ~ \langle\Psi |~ |\Phi \rangle $ provided we
perform complex conjugation of the weak value together with the
interchange. This is similar to complex conjugation of the Schr\"odinger
wave function under the time reversal. Thus, also for weak measurements
there is the time reversal symmetry: both  $\langle\Phi |$ and  $|\Psi
\rangle$ enter the formula of the weak value on the same footing and
there is symmetry under the interchange of the pre- and post-selected states.
 The time symmetry holds for weak values of generalized two-state
 vectors (\ref{wv-gen}):
i.e., the interchange \break
$\sum_i \alpha_i \langle \Phi_i | ~ | \Psi_i \rangle  ~ \leftrightarrow
~ \sum_i \alpha_i^{\ast} \langle \Psi_i | ~ | \Phi_i \rangle~$ leads to
$~C_w  \leftrightarrow C_w^\ast$.

\subsection{The time-asymmetry}
\label{sec:tas}

 The symmetry is also suggested in using the language of
 ``pre-selected'' state and ``post-selected'' state.  In order to
 obtain the two-state vector (\ref{2sv}) we need to pre-select $A=a$ at $t_1$
 and post-select $B=b$ at $t_2$. Both measurements might not yield the
 desired outcomes, so we need several systems out of which we pre- and
 post-select the one which is described by the two-state vector
 (\ref{2sv}). However, the symmetry is not complete and the language
 might be somewhat misleading. It is true that we can only
 (post-)select $B=b$ at $t_2$, but we can {\it prepare} instead of
 pre-select $A=a$ at $t_1$.  For preparation of $|a\rangle$ a single
 system is enough. If the measurement of $A$ yields a different
 outcome $a'$ we can perform a fast unitary operation which will
 change $|A=a'\rangle$ to $|A=a\rangle$ and then the time evolution to
 time $t$ will bring the system to the state $ |\Psi \rangle $. This
 procedure is impossible for creation of the backward evolving
 state $\langle\Phi |$. Indeed, if the outcome of the measurement of
 $B$ does not yield $b$, we cannot read it and then make an
 appropriate unitary operation {\it before} $t_2$ in order to get the
 state $\langle\Phi |$ at time $t$. We need several systems to
 post-select the desired result (unless by chance the first system has
 the desired outcome).

 Although the formalism includes situations with descriptions by solely forward
 evolving quantum state and by solely backward-evolving quantum states,
  here also there is a conceptual difference. For
  obtaining backward evolving state it was necessary to have a
  guarded ancilla in order to erase the quantum state evolving from
  the past. Of course, there is no need for this complication in obtaining
  forward-evolving quantum state. The difference is due to fixed
  ``memory'' arrow of time: we know the past and we do not know the
  future. This asymmetry is also connected to the concept of a
  measurement.  It is asymmetric because, by definition, we do not know
  the measured value before the measurement and we do know it after
  the measurement.

\subsection{If {\it measurements} are time-asymmetric, how the
  outcomes of measurements are time-symmetric? }
\label{sec:syasy}

Taking this asymmetry of the concept of measurement  into account, how one can
understand the time symmetry of the formulae for  the probability of
an intermediate measurements (\ref{ABL}), (\ref{ABL-gen}) and for the
formulae of weak values  (\ref{wv}), (\ref{wv-gen})?

This is because these formulae deal with the results of the
measurements which, in contrast with the concept of measurement
itself, are free from the time-asymmetry of a measurement. The results
of measurements represent the way the system affects other systems (in
this case  measuring devices) and these effects, obviously, do not
exhibit the time asymmetry of our memory. The time
asymmetry of measurement is due to the fact that the pointer variable
of the measuring device is showing ``zero'' mark before the measurement
and not after the measurement. But the result of the measurement is
represented by the {\it shift} of the pointer position. (If
originally the pointer showed ``zero'' it is also represented by the
final position of the pointer.) This shift is independent of the
initial position of the pointer and therefore it is not sensitive to
the time-asymmetry caused by asymmetrical fixing of the initial (and
not final) position of the pointer. The relations described in the
 formulae of the two-state vector formalism are related to these
shifts and, therefore, the time-symmetry of the formulae follows from the
underlying time symmetry of the quantum theory.
The shifts of the pointer variable in weak measurements were
considered as ``weak-measurements elements of reality'' \cite{WMER}
where ``elements of reality were identified with ``definite
shifts''. This approach was inspired by the EPR elements of reality
which are definite outcomes of ideal measurements, i.e., definite
shifts in ideal measurement procedures. The next section discusses a
controversy related to ideal measurements.

\subsection{ Counterfactual Interpretation of the ABL Rule~~}
\vskip .2 cm

Several authors  criticized the TSVF because of the alleged conflict
between counterfactual interpretations of the ABL rule and
predictions of quantum theory \cite{SS,CO,MI,Kastold}. The  form of
all these inconsistency proofs is as follows: The probability of an
outcome $C=c_n$ of a quantum measurement performed on a pre-selected
system, given correctly by (\ref{ABL-pre}), is considered. In order
to allow the analysis using the ABL formula, a measurement at a
later time, $t_2$, with two possible outcomes, which we denote by
``$1_f$'' and ``$2_f$'',
 is introduced.
The suggested application of the ABL rule is expressed in the formula
for the
probability of the result $C=c_n$
\begin{eqnarray}
  \label{p1-lev}
  {\rm Prob}(C=c_n)&=&  {\rm Prob}(1_f) ~ {\rm Prob}(C=c_n~; 1_f)
\\
\nonumber
  &+&
  {\rm Prob}(2_f) ~{\rm Prob}(C=c_n ~; 2_f),
\end{eqnarray}
where $ {\rm Prob}(C=c_n ~; 1_f)$ and $ {\rm Prob}(C=c_n~ ; 2_f)$ are the
conditional probabilities given by the ABL formula, (\ref{ABL}), and $ {\rm
  Prob}(1_f)$ and $ {\rm Prob}(2_f)$ are the probabilities of the
results of the final measurement. In the proofs, the authors show
that Eq.~(\ref{p1-lev}) is not valid and conclude that  the ABL
formula is not applicable to this example and therefore  it is not
applicable in general.

One us (LV) has  argued \cite{DTSQT,TSCF,CO-co} that the error in
calculating equality (\ref{p1-lev}) does not arise from the
conditional probabilities given by the ABL formula, but from the
calculation of the probabilities $ {\rm Prob}(1_f)$ and $ {\rm
Prob}(2_f)$ of the final measurement. In all three alleged proofs
the probabilities $ {\rm
  Prob}(1_f)$ and $ {\rm Prob}(2_f)$ were calculated on the assumption
that {\rm no} measurement took place at time $t$. Clearly, one
cannot make this assumption here since then the discussion about the
probability of the result of the measurement at time $t$ is
meaningless. Thus, it is not surprising that the value of the
probability ${\rm Prob}(C=c_n)$ obtained in this way comes out
different from the value predicted by the quantum theory.
Straightforward calculations show that the formula (\ref{p1-lev})
with the probabilities $ {\rm Prob}(1_f)$ and $ {\rm Prob}(2_f)$
calculated on the condition that the intermediate measurement has
been performed leads to the result predicted by the standard
formalism of quantum theory.

The analysis of counterfactual statements considers both {\em actual}
and {\em counterfactual} worlds. The statement is considered to be
true if it is true in counterfactual worlds ``closest'' to the actual
world.  In the context of the ABL formula, in the actual world the
pre-selection and the post-selection has been successfully performed,
but the measurement of $C$ has not (necessarily) been performed. On
the other hand, in counterfactual worlds the measurement of $C$ has
been performed.  The problem is to find counterfactual worlds
``closest'' to the actual world in which the measurement of $C$ has
been performed. The fallacy in all the inconsistency proofs is that
their authors have considered counterfactual worlds in which $C$ has
not been measured.

Even if we disregard  this fallacy  there is still a difficulty in defining the
``closest'' worlds in the framework of the TSVF. In standard quantum
theory it is possible to use the most natural definition of the
``closest'' world. Since the future is considered to be irrelevant for
measurements at present time $t$, only the period of time before $t$
is considered. Then the definition is:
\begin{quotation}
{\bf (i)} Closest counterfactual  worlds are  the
worlds in which the system is described by the same quantum state as
in the actual world.
 \end{quotation}
In the framework of the TSVF, however, this definition is not
acceptable. In the time-symmetric approach the period of time before
and after $t$  is considered. The measurement of $C$ constrains the
possible states immediately after $t$ to the eigenstates of $C$.
Therefore,  if in the actual world the state immediately after $t$
is  not an eigenstate of $C$, no counterfactual world with the same
state exists. Moreover, there is the same problem with the backward
evolving quantum state (the concept which does not exist in the
standard approach) in the period of time before  $t$. This
difficulty can be solved by adopting the following definition of the
closest world \cite{TSCF}:
\begin{quotation}
{\bf (ii)}  Closest counterfactual  worlds are  the
worlds in which the results of all measurements performed on the system
(except the measurement at time $t$) are the same as in
the  actual world.
 \end{quotation}
 For the
 pre-selected only situation, this definition is equivalent to  (i),
 but it is also applicable to the
symmetric pre- and post-selected situation. The definition allows to
construct {\it time-symmetric counterfactuals} in spite of common
claims that such concept is inconsistent \cite{Lewis}.

An important example of  counterfactuals in quantum
theory are  ``elements of reality'' which are inspired by the EPR
elements of reality.
The  modification of the definition of elements of reality  applicable
  to the framework of the TSVF \cite{PR} is:
 \begin{quotation}
 {\bf (iii)~}   If we can {\em infer} with certainty that the result of
   measuring at time $t$ of an observable $C$ is $c$, then, at time
   $t$, there exists an element of reality $C=c$.
\end{quotation}
The word ``infer'' is neutral relative to past and future. The
inference about results at time $t$ is based on the results of
measurements on the system performed both before and after time $t$.
Note that there are situations (e.g., the three-boxes example) in
which we can ``infer'' some facts that cannot be obtained by neither
``prediction'' based on the past results nor ``retrodiction'' based on
the future results separately.

The theorem (i) of Sec. \ref{sec:relation} now can be formulated in a
simple way: If $A=a$ is an element of reality then $A_w =a$ is the
weak-measurement of reality. The theorem (ii) of
Sec. \ref{sec:relation} can be formulated as follows. If $A$ is a
dichotomic variable, $a$ is an eigenvalue of $A$, and if   $A_w =a$
is a weak-measurement element of reality,
then  $A=a$ is an element of reality.

The discussion about the meaning of time symmetric counterfactuals
continues until today. Kastner changed her view on such
counterfactuals from ``inconsistent'' to ``trivial'' \cite{KastPS}.
See Vaidman's
 reply  \cite{myrepKast} and other very recent contributions on this
 issue \cite{Kast04,LeSp,Mi06}.

\section{Protective measurements}

Several years ago we proposed a concept of {\it protective
  measurements} \cite{PM1,PM2,PM3} which provides an argument strengthening
the consideration of a quantum state as a ``reality'' of some kind.
We have shown that ``protected'' quantum states can be observed just
on a single quantum system. On the other hand, if a single quantum
state is ``the reality'' how ``the two-state vector'' can be ``the
reality''?

\subsection{ Protective measurement of a single quantum state}

In order to measure the quantum state of single system one has to
measure expectation values of various observables.
In general, the weak (expectation) value cannot be measured on a
single system. However, it can be done if the quantum state is {\em
protected} \cite{PM1,PM2}. The appropriate measurement interaction
is again described  the Hamiltonian (\ref{neumann}),
but instead of an impulsive
interaction the adiabatic limit of slow and weak interaction is
considered:
$g(t) = 1/T$ for most of the interaction time $T$ and
 $g(t)$ goes to zero gradually before and after the period
$T$.

In this case the interaction Hamiltonian  does not dominate the
time evolution during the measurement, moreover, it can be considered
as a perturbation. The free Hamiltonian $H_0$ dominates the evolution.
In order to protect a  quantum
state this Hamiltonian must have the state to be a non-degenerate
energy eigenstate.
For $g(t)$ smooth enough we then obtain an adiabatic process in
which the system cannot make a transition from one energy eigenstate
to another, and, in the limit $T \rightarrow \infty$, the interaction
Hamiltonian          changes
 the energy eigenstate by an infinitesimal amount.
   If the initial state
of the system is an eigenstate $\vert E_i\rangle$
of $H_0$ then for any given value of
$P$, the energy of the eigenstate shifts by an infinitesimal amount
given by the first order perturbation theory:
$\delta E = \langle E_i \vert H_{int} \vert E_i  \rangle  =
\langle E_i \vert
A \vert E_i\rangle P/ T.
$ The corresponding time evolution $ e^{-i P \langle E_i \vert
A \vert E_i\rangle } $ shifts the
pointer by the expectation  value of
$ A$ in the state $\vert E_i\rangle$.
Thus, the probability distribution of the pointer variable,
$ e^{ -{{(Q-a_i)^2} /{\Delta ^2}}}$ remains unchanged in
its shape, and  is shifted by the
 expectation value
$\langle A \rangle_i=\langle E_i\vert A\vert E_i\rangle$.

If the initial state of the system is a superposition of several
non-degenerate energy eigenstates
$
|\Psi_1 \rangle = \Sigma \alpha_i |E_i \rangle
$,
then
a particular outcome $\langle A \rangle_i \equiv \langle E_i \vert A
\vert E_i\rangle$
appears at random, with the probability $|\alpha_i|^2$
\cite {Unruh}.
(Subsequent adiabatic measurements of the same observable $A$ invariably yield
the expectation value in the same eigenstate $\vert E_i\rangle$.)

 \subsection{Protective Measurement of a Two-State Vector}

 At first sight, it seems that protection of a two-state
vector is impossible. Indeed, if we add a potential that makes one
state  a non-degenerate eigenstate, then the other state, if it is
different, cannot be an eigenstate too. (The states of the two-state
vector cannot be orthogonal.)  But, nevertheless, protection of the
two-state vector is possible \cite{PM2sv}.

 The procedure for protection of a two-state vector of a given system is
accomplished by coupling the system to another  pre- and post-selected
system. The protection procedure takes advantage of the fact that
weak values  might  acquire complex values. Thus, the effective
Hamiltonian of the protection might not be Hermitian. Non-Hermitian
Hamiltonians act  in different ways on quantum states evolving forward and
backwards in time. This allows simultaneous protection of two different
states (evolving in opposite time directions).

Let us consider  an example \cite{PM2sv} of a  two-state
vector of a spin-$1\over2$ particle, $\langle{\uparrow_y}
|  |{\uparrow_x} \rangle $.
The protection procedure uses an external   pre- and post-selected
system $S$ of a large
spin $N$ that is coupled to our spin via the interaction
\begin{equation}
  \label{hprot}
H_{prot} = - \lambda {\bf S \cdot \sigma}.
\end{equation}
The external system is pre-selected in the state $|S_x {=} N\rangle$ and
post-selected in the state $\langle S_y {=} N|$, that is, it is described by
the two-state
vector $\langle S_y {=} N
|  |S_x {=} N \rangle $. The coupling constant $\lambda$ is chosen in such a
way that the
interaction with our spin-$1\over 2$ particle  cannot
change significantly the two-state vector of the protective system $S$, and
the spin-$1\over 2$ particle ``feels'' the effective Hamiltonian in which
{\bf $S$} is
replaced by its weak value,
\begin{equation}
  \label{sweak}
{\bf S}_w = {{\langle S_y = N
|(S_x, S_y, S_z)  |S_x = N  \rangle} \over{\langle S_y = N
 |S_x = N  \rangle}} = (N, N, iN)  .
\end{equation}
Thus, the effective protective Hamiltonian is
\begin{equation}
  \label{heff}
H_{eff} =- \lambda N( \sigma_x +  \sigma_y + i\sigma_z).
\end{equation}
The state $|{\uparrow_x} \rangle $ is an eigenstates of this
(non-Hermitian) Hamiltonian (with eigenvalue $-\lambda N$).  For
backward evolving states the effective Hamiltonian is the hermitian
conjugate of (\ref{heff}) and it has different (non-degenerate) eigenstate with
this eigenvalue; the eigenstate is $\langle {\uparrow_y}|$.

 In order to prove that the Hamiltonian (\ref{hprot}) indeed provides the
protection, we have to show that the two-state vector $\langle{\uparrow_y}
|   |{\uparrow_x} \rangle $ will remain essentially unchanged during the
measurement.
We consider measurement which is performed during the period of time,
between pre- and post-selection which we choose to be equal one.
The Hamiltonian
\begin{equation}
  \label{ham}
H = -\lambda {\bf S \cdot \sigma} + P  \sigma_{\xi}
\end{equation}
can be replaced by the effective Hamiltonian
\begin{equation}
  \label{ham-eff}
H_{eff} =- \lambda N( \sigma_x +  \sigma_y + i\sigma_z) + P  \sigma_{\xi} .
\end{equation}
Indeed, the system with the spin $S$ can be considered as $N$ spin 1/2
particles all pre-selected in $|{\uparrow_x} \rangle$ state and
post-selected in $|{\uparrow_y} \rangle$ state. The strength of the
coupling to each spin 1/2 particle is $\lambda \ll 1$, therefore
during the time of the measurement their states cannot be changed
significantly. Thus, the forward evolving state
$|S_x {=} N\rangle$ and  the  backward evolving state $\langle S_y {=}
N|$ do not change significantly during the measuring process. The
effective coupling to such system is the coupling to its weak values.

Good precision of the measurement of the spin component requires large
uncertainty in $P$, but we can arrange the experiment in such a way
that $P \ll \lambda N$. Then the second term in the Hamiltonian \ref{ham-eff}  will not
change significantly the eigenvectors. The two-state vector
$\langle{\uparrow_y} | |{\uparrow_x} \rangle $ will remain essentially
unchanged during the measurement, and therefore the measuring device
on this single particle will yield $({\sigma_\xi})_w =
{{\langle{\uparrow_y} |\sigma_\xi |{\uparrow_x} \rangle}\over
{\langle{\uparrow_y} |{\uparrow_x} \rangle}}$.

 The Hamiltonian (\ref{ham}), with an  external system described
by the two-state vector $\langle S_y = N
|  |S_x = N \rangle $, provides protection for the two-state vector
$\langle{\uparrow_y}|  |{\uparrow_x} \rangle $. It is not difficult to
demonstrate
that any two-state vector obtained by pre- and post-selection of  the
spin-$1\over 2$ particle can be protected by the Hamiltonian (\ref{ham}). A general form of
the two-state vector is $\langle{\uparrow_\beta}|  |{\uparrow_\alpha}
\rangle $ where  $\hat{\alpha}$ and $\hat{\beta}$ denote some directions. It can be
verified by a straightforward calculation that the two-state vector $\langle{\uparrow_\beta}|  |{\uparrow_\alpha}
\rangle $ is protected when  the two-state vector of the protective device
is  $\langle S_\beta = N
|  |S_\alpha = N \rangle $.

At least formally we can generalize this method to make a protective
measurement of an arbitrary two-state vector $\langle \Psi _2|  | \Psi
_1 \rangle $ of an arbitrary system.
 Let us decompose the post-selected state
 $| \Psi _2 \rangle = a | \Psi
_1 \rangle + b | \Psi _\bot \rangle $.
 Now we can define ``model spin'' states:
$ | \Psi _1 \rangle \equiv | \tilde{\uparrow}_z \rangle$ and
$ | \Psi _\bot \rangle
\equiv | \tilde{\downarrow}_z \rangle$.
 On the basis of the two orthogonal states we
can obtain all other ``model spin'' states. For example,
$ |\tilde{\uparrow}_x \rangle = 1/\sqrt 2 ~( |\tilde{\uparrow}_z \rangle +
|\tilde{\downarrow}_z \rangle)$,
and then we can define
 the ``spin model'' operator $\bf \tilde{\sigma}$. Now, the
protection Hamiltonian,
in complete analogy with the spin-$1\over 2$ particle case
is
\begin{equation}
  \label{ham-eff-pr}
H_{prot} = - \lambda {\bf S \cdot \tilde{\sigma}}.
\end{equation}
In order to protect the state $\langle \Psi _2|  | \Psi _1 \rangle $, the
pre-selected state of the external system has to be $|S_z {=} N \rangle$
and the
post-selected state has to be  $\langle S_\chi {=} N|$ where the direction
$\hat{\chi}$ is defined by the  ``spin model'' representation of the state
$| \Psi _2\rangle$,
\begin{equation}
  \label{ham-mod}
|\tilde{\uparrow}_\chi \rangle \equiv | \Psi _2\rangle = \langle \Psi _1
| \Psi _2 \rangle|\tilde{\uparrow}_z \rangle + \langle \Psi _\bot
| \Psi _2 \rangle|\tilde{\downarrow}_z \rangle .
\end{equation}

However, this scheme usually leads to unphysical interaction and is
good only as a gedanken experiment in the framework of non-relativistic
quantum theory where we assume that any hermitian Hamiltonian is possible.

\section{The TSVF and the Many-Worlds Interpretation of quantum theory}
\vskip .2cm

The TSVF fits very well into the many-worlds interpretation (MWI)
\cite{MWI}, the preferred interpretation of quantum theory of one of
us (LV) \cite{my-MWI}.  The counterfactual worlds corresponding to
different outcomes of quantum measurements have in the MWI an
especially clear meaning: these are subjectively actual different
worlds.  In each world the observers of the quantum measurement call
their world the actual one, but, if they believe in the MWI they
have no paradoxes about ontology of the other worlds. The apparent
paradox that a weak value at a given time might change from an
expectation value to a weak value corresponding to a particular
post-selection is solved in a natural way: in a world with
pre-selection only (before the post-selection) the weak value is the
expectation value; then this world splits into several worlds
according to results of the post-selection measurement and in each
of these worlds the weak value will be that corresponding to the
particular post-selection.  The time-symmetric concepts of
``elements of reality'', ``weak-measurements elements of reality''
are consistent and meaningful in the context of a particular world.
Otherwise, at time $t$, before the ``future'' measurements have been
performed,  the only meaningful concepts are the concepts of the
standard, time-asymmetric approach.

One of us (YA) is not ready to adopt the far reaching consequences of
the MWI. He proposes another solution. It takes the TSVF even more
seriously than it was presented in this paper. Even at present, before
the ``future'' measurements, the backward evolving quantum state (or
its complex conjugate evolving forward in time) exists! It exists in
the same way as the quantum state evolving from the past exists. This
state corresponds to particular outcomes of all measurements in the
future. An element of arbitrariness: ``Why this particular outcome and
not some other?'' might discourage, but the alternative (without the
many-worlds) -- the collapse of the quantum wave -- is clearly worse than that.

\vskip .4cm \noindent {\bf Acknowledgments}

It is a pleasure to thank Fred Alan Wolf for correcting an error of
the previous version of the review and  Juan Gonzalo Muga for
helpful suggestions. This work has been supported in part by the
European Commission under the Integrated Project Qubit Applications
(QAP) funded by the IST directorate as Contract Number 015848 and by
grant 990/06 of the Israel Science Foundation.

\end{document}